\documentclass[aps,onecolumn,letterpaper,oneside,preprint,tightenlines,draft,%
eqsecnum,nobibnotes,nofootinbib,amsfonts,amssymb,amsmath,showpacs,%
showkeys]{revtex4}

\allowdisplaybreaks

\newcommand{\cN}{\mathcal{N}}

\newcommand{\bH}{\boldsymbol{H}}
\newcommand{\bQ}{\boldsymbol{Q}}
\newcommand{\fF}{\mathfrak{F}}
\newcommand{\hA}{\Hat{A}}
\newcommand{\hE}{\Hat{E}}
\newcommand{\hQ}{\Hat{Q}}
\newcommand{\hW}{\Hat{W}}
\newcommand{\bbR}{\mathbb{R}}
\newcommand{\bbC}{\mathbb{C}}
\newcommand{\rmi}{\mathrm{i}}
\newcommand{\rme}{\mathrm{e}}
\newcommand{\rmd}{\mathrm{d}}
\newcommand{\del}{\partial}

\newcommand{\lra}{\leftrightarrow}
\newcommand{\rnu}{\sqrt{\nu}}
\newcommand{\rmj}{\mathrm{j}}
\newcommand{\rmP}{\mathrm{P}}
\newcommand{\sV}{\mathsf{V}}

\begin{document}


%
%

\title{Existence of Different Intermediate Hamiltonians in Type A
 $\cN$-fold Supersymmetry II. The $\cN=3$ Case}
\author{Bijan Bagchi}
\email{bbagchi123@rediffmail.com}
\affiliation{Department of Applied Mathematics, University of
 Calcutta,\\
 92 Acharya Prafulla Chandra Road, Kolkata 700 009, India} 
\author{Toshiaki Tanaka}
\email{ttanaka@mail.ncku.edu.tw}
\affiliation{Department of Physics, National Cheng Kung University,\\
 Tainan 701, Taiwan, R.O.C.\\
 National Center for Theoretical Sciences, Taiwan, R.O.C.}


\begin{abstract}

We continue the previous study on the existence of different
intermediate Hamiltonians in type A $\cN$-fold supersymmetric
systems and carry out an exhaustive investigation on the $\cN=3$
case. In contrast with the $\cN=2$ case, we find various patterns
in the existence of intermediate Hamiltonians due to the presence
of two different intermediate positions in a factorized type A
$3$-fold supercharge. In addition, all the $\cN=3$ models are
strictly restricted to at most elliptic type, which enables us
to make the complete classification of the systems which admit
intermediate Hamiltonians. Finally, we show realizations of
third-order parasupersymmetry and variant generalized $3$-fold
superalgebras by such systems.

\end{abstract}


\pacs{03.65.Ca; 03.65.Fd; 11.30.Na; 11.30.Pb}
\keywords{$\cN$-fold supersymmetry; Parasupersymmetry; Intertwining
 operators; Quasi-solvability; Shape invariance; Generalized
 superalgebra}




\maketitle

\section{Introduction}
\label{sec:intro}

Recently we reported~\cite{BT09} on the existence of different
intermediate Hamiltonians in type A $\cN$-fold supersymmetry
(SUSY)~\cite{AST01a}.
Apart from the connectivity to a wide range of solvable and
quasi-solvable potentials including not only the well-known
$\mathfrak{sl}(2)$ Lie-algebraic models~\cite{Tu88} but also the new
extended class of completely solvable rational ones~\cite{BQR09},
type A $\cN$-fold SUSY is of independent interest
because of the rich mathematical structure it provides (for a review,
see Ref.~\cite{Ta09}). 

The main motivation of our study came from the issue of non-uniqueness
of factorizing operators as a consequence of the underlying $GL(2,\bbC)$
symmetry. Type A $\cN$-fold supercharge admits of a one-parameter
family of factorizations that is expressible as a product of $\cN$
first-order linear differential operators due to the aforementioned
symmetry~\cite{Ta03a}. This in turn implies that a type A $\cN$-fold
SUSY can have different intermediate Hamiltonians corresponding to
different factorizations. However the existence of intermediate
Hamiltonians is not guaranteed in general~\cite{AST01a}. In
Ref.~\cite{BT09}, we investigated under what conditions type A
$\cN$-fold SUSY systems can admit intermediate Hamiltonians and
that how many sets of such Hamiltonians are plausible for the specific
case of $\cN=2$. We then concluded that the number of admissible
intermediate Hamiltonians would be a more suitable index than the
concept of \emph{reducibility} introduced in Refs.~\cite{AICD95,AIN95a}
to characterize higher-order intertwining operators. Furthermore,
we found that it naturally leads to a realization of second-order
parasupersymmetry (paraSUSY)~\cite{RS88} and generalized $2$-fold
superalgebra~\cite{Ta07a}.

In this article, we pursue our investigations further and focus on the
$\cN=3$ case. In comparison with the previous $\cN=2$ case, there appear
mainly two novel features in the $\cN=3$ case. The one is the fact that
the functional types of type A $\cN$-fold SUSY potentials for $\cN\geq3$
are strictly restricted to at most elliptic functions due to an
additional constraint. But due to this constraint, all the type A
$\cN$-fold SUSY models for $\cN\geq3$ were completely classified in
Ref.~\cite{Ta03a}. We find that we can also classify entirely all
the type A $3$-fold SUSY models which have intermediate Hamiltonians.

The other novel feature of $\cN=3$ is concerned with the fact that
there are two different intermediate positions in each
factorized $3$-fold supercharge. Due to the latter fact, one
can consider different patterns in the existence of intermediate
Hamiltonians. That is, in certain cases systems admit intermediate
Hamiltonians at both the two intermediate positions while in other
cases systems admit them at only one of the two positions.
Explorations of the underlying conditions for each case reveal that for
the former classes a system with intermediate Hamiltonians at the two
positions turns out to be always solvable and even shape invariant.
On the other hand, for the latter classes a system with an intermediate
Hamiltonian at only one position is led to quasi-solvability only.

Thanks to the aforementioned different patterns in the $\cN=3$ case,
we further find intriguing and quite rich structure which does not
exist in the $\cN=2$ case when we consider the existence of more than
one sets of intermediate Hamiltonians. For instance, we find that
there are systems which have intermediate Hamiltonians at the two
positions in one factorization but has only one at one of the
positions in another factorization. Throughout the analyses of such
systems, we realize that in contrast to the $\cN=2$ case we must
consider not only the number of admissible intermediate Hamiltonians
but also the variety in the existence of them to characterize
$\cN$th-order intertwining operators for the $\cN\geq3$ cases.

We also investigate a parafermionic formulation of such systems and
realizations of third-order paraSUSY~\cite{To92,Kh92} and generalized
$3$-fold superalgebra~\cite{Ta07c}. We find not only that such
realizations are indeed possible but also that variations of the
latter superalgebra hold according to the different patterns in the
existence of intermediate Hamiltonians.

The article is organized as follows.
In Section~\ref{sec:A3S}, we review type A 3-fold SUSY and discuss
some of its salient features. Then, we introduce three classes
according to three different patterns in the existence of intermediate
Hamiltonians. In Section~\ref{sec:exinh}, we work out explicitly
the necessary and sufficient conditions for each of the three classes.
In Section~\ref{sec:cls1}, we classify exhaustively different
type A 3-fold SUSY potentials with one set of intermediate
Hamiltonians. In particular, we find an intimate relation between
the patterns and the degree of solvability. That is,
all the systems which have intermediate Hamiltonians at the two
positions consist of shape invariant potentials. In contrast, those
which have one at only one position comprise sextic anharmonic
oscillators, deformed P\"{o}schl-Teller or Scarf potentials, and
one-body elliptic Inozemtsev potentials, all of which are
quasi-solvable. In Section~\ref{sec:exdif}, we discuss and study in
full detail the necessary and sufficient conditions for a system to
have different sets of intermediate Hamiltonians in various patterns.
In Section~\ref{sec:cls2}, we then give the complete classification
of the type A $3$-fold SUSY models which admit simultaneously more
than one sets of intermediate Hamiltonians. In Section~\ref{sec:qpara},
we show that a system with intermediate Hamiltonians at the two
positions always admits a realization of third-order paraSUSY. In
addition, each system belonging to one of the three classes always
admits a realization of variant generalized $3$-fold superalgebras.
In the final section, we summarize and discuss various aspects of
the obtained results.

\section{Type A 3-fold Supersymmetry}
\label{sec:A3S}

A type A 3-fold SUSY system is given by
\begin{subequations}
\label{eqs:A3sys}
\begin{gather}
H^{\pm}=-\frac{1}{2}\frac{\rmd^{2}}{\rmd x^{2}}+\frac{1}{2}W(x)^{2}
 -\frac{1}{3}\left(2E'(x)-E(x)^{2}\right)-R\pm\frac{3}{2}W'(x),
\label{eq:A3Ham}\\
P_{3}^{-}=P_{31}^{-}P_{32}^{-}P_{33}^{-},\qquad P_{3}^{+}=-(P_{3}^{-}
 )^{\text{T}}=P_{33}^{+}P_{32}^{+}P_{31}^{+},
\end{gather}
\end{subequations}
where the superscript T denotes the transposition in the $x$-space
and $P_{3i}^{\pm}$ ($i=1,2,3$) are defined by
\begin{align}
P_{31}^{\pm}=\mp\del+W-E,\qquad P_{32}^{\pm}=\mp\del+W,\qquad
 P_{33}^{\pm}=\mp\del+W+E.
\label{eq:P3i+-}
\end{align}
In the expanded form, the type A 3-fold supercharge component
$P_{3}^{-}$ reads
\begin{align}
P_{3}^{-}=&\:\del^{3}+3W\del^{2}
 +\left(3W'+3W^{2}+2E'-E^{2}\right)\del\notag\\
&+W''+3WW'+W^{3}+\left(2E'-E^{2}\right)W
 +\frac{1}{2}\left(2E'-E^{2}\right)'.
\label{eq:P3-}
\end{align}
The functions $E(x)$ and $W(x)$ are not arbitrary but are connected
with a fourth-degree polynomial $A(z)$ and a second-degree polynomial
$Q(z)$
\begin{subequations}
\label{eqs:polAQ}
\begin{align}
A(z)&=a_{4}z^{4}+a_{3}z^{3}+a_{2}z^{2}+a_{1}z+a_{0},\\
Q(z)&=b_{2}z^{2}+b_{1}z+b_{0},
\end{align}
\end{subequations}
through the following relations
\begin{align}
z''(x)=E(x)z'(x),\quad 2A(z)=z'(x)^{2},\quad Q(z)=-z'(x)W(x).
\label{eq:defAQ}
\end{align}
{}From these relations, we obtain in particular
\begin{align}
W'(x)+E(x)W(x)=-Q'(z),\qquad E'(x)+E(x)^{2}=A''(z),
\label{eq:Q'A''}
\end{align}
which will play important roles in later analyses. The restriction
on the form of $A(z)$ arises in type A $\cN$-fold SUSY for all
$\cN\geq3$ and is absent for $\cN=2$. This strongly limits the possible
forms of potentials for $\cN\geq3$.

Due to the underlying algebraic structure of type A $\cN$-fold
SUSY systems, they are invariant under the linear projective
transformations of $GL(2,\bbC)$ defined by
\begin{align}
z=\frac{\alpha w+\beta}{\gamma w+\delta},\quad
 (\alpha,\beta\gamma,\delta\in\bbC,\ \Delta=\alpha\delta
 -\beta\gamma\neq0).
\end{align}
The polynomials $A(z)$ and $Q(z)$ are then transformed under
the $GL(2,\bbC)$ transformations as
\begin{align}
A(z)\mapsto\hA(w)=\Delta^{-2}(\gamma w+\delta)^{4}A(z)\Bigr|_{z=
 \frac{\alpha w+\beta}{\gamma w+\delta}}=\sum_{k=0}^{4}
 \hat{a}_{k}w^{k},\\
Q(z)\mapsto\hQ(w)=\Delta^{-1}(\gamma w+\delta)^{2}Q(z)\Bigr|_{z=
 \frac{\alpha w+\beta}{\gamma w+\delta}}=\sum_{k=0}^{2}
 \hat{b}_{k}w^{k},
\end{align}
where the new coefficients $\hat{a}_{i}$ ($i=0,\dots,4$) and
$\hat{b}_{i}$ ($i=0,1,2$) are respectively given by
\begin{multline}
\left(
 \begin{array}{r}
 \hat{a}_{4}\\ \hat{a}_{3}\\ \hat{a}_{2}\\ \hat{a}_{1}\\ \hat{a}_{0}
 \end{array}
\right)=\Delta^{-2}\left(
 \begin{array}{rr}
 \alpha^{4} & \alpha^{3}\gamma\\
 4\alpha^{3}\beta & \alpha^{2}(\alpha\delta+3\beta\gamma)\\
 6\alpha^{2}\beta^{2} & 3\alpha\beta(\alpha\delta+\beta\gamma)\\
 4\alpha\beta^{3} & \beta^{2}(3\alpha\delta+\beta\gamma)\\
 \beta^{4} & \beta^{3}\delta\\
 \end{array}\right.\\
\left.
 \begin{array}{rrr}
 \alpha^{2}\gamma^{2} & \alpha\gamma^{3} & \gamma^{4}\\
 2\alpha\gamma(\alpha\delta+\beta\gamma) &
  \gamma^{2}(3\alpha\delta+\beta\gamma) & 4\gamma^{3}\delta\\
 \alpha^{2}\delta^{2}+4\alpha\beta\gamma\delta+\beta^{2}\gamma^{2}
  & 3\gamma\delta(\alpha\delta+\beta\gamma) & 6\gamma^{2}\delta^{2}\\
 2\beta\delta(\alpha\delta+\beta\gamma) &
  \delta^{2}(\alpha\delta+3\beta\gamma) & 4\gamma\delta^{3}\\
 \beta^{2}\delta^{2} & \beta\delta^{3} & \delta^{4}
 \end{array}
\right)\left(
 \begin{array}{r}
 a_{4}\\ a_{3}\\ a_{2}\\ a_{1}\\ a_{0}
 \end{array}
\right),
\label{eq:tfas}
\end{multline}
and
\begin{align}
\left(
 \begin{array}{r}
 \hat{b}_{2}\\ \hat{b}_{1}\\ \hat{b}_{0}
 \end{array}
\right)=\Delta^{-1}\left(
 \begin{array}{rrr}
 \alpha^{2} & \alpha\gamma & \gamma^{2}\\
 2\alpha\beta & \alpha\delta+\beta\gamma & 2\gamma\delta\\
 \beta^{2} & \beta\delta & \delta^{2}
 \end{array}
\right)\left(
 \begin{array}{r}
 b_{2}\\ b_{1}\\ b_{0}
 \end{array}
\right).
\label{eq:tfbs}
\end{align}
{}From the formulas (\ref{eq:defAQ}), the induced transformations
of functions $E(x)$ and $W(x)$ read
\begin{align}
W(x)\mapsto\hW(x)=W(x),\qquad
 E(x)\mapsto\hE(x)=E(x)-\frac{2\gamma z'(x)}{\gamma z(x)-\alpha}.
\label{eq:tfEW}
\end{align}
The latter transformation rule in particular implies the invariance
of the following function:
\begin{align}
2\hE'(x)-\hE(x)^{2}=2E'(x)-E(x)^{2}.
\end{align}
The $GL(2,\bbC)$ invariance of any type A 3-fold SUSY system is
now manifest since both the pair of Hamiltonians $H^{\pm}$ in
(\ref{eq:A3Ham}) and the operator $P_{3}^{-}$ in (\ref{eq:P3-}) only
depend on $W$ and $2E'-E^{2}$ which are both invariant under any
$GL(2,\bbC)$ transformations.

The superHamiltonian $\bH_{3}$ and the type A $3$-fold supercharges
$\bQ_{3}^{\pm}$ introduced with the ordinary fermionic variables
$\psi^{\pm}$ as
\begin{align}
\bH_{3}=H^{-}\psi^{-}\psi^{+}+H^{+}\psi^{+}\psi^{-},\qquad
 \bQ_{3}^{\pm}=P_{3}^{\mp}\psi^{\pm},
\label{eq:ofrep}
\end{align}
satisfy the type A $3$-fold superalgebra~\cite{Ta03a}
\begin{subequations}
\label{eqs:A3alg}
\begin{align}
\bigl[\bQ_{3}^{\pm},\bH_{3}\bigr]=&\:\bigl\{\bQ_{3}^{-},\bQ_{3}^{+}
 \bigr\}=0,
\label{eq:A3alg1}\\
\bigl\{\bQ_{3}^{-},\bQ_{3}^{+}\bigr\}=&\:8(\bH_{3}+R)^{3}-\frac{8}{3}
 \left(i_{2}[A]-3D_{2}[Q]\right)(\bH_{3}+R)\notag\\
&+\frac{16}{27}\left(j_{3}[A]+9I_{1,2}[A,Q]\right),
\label{eq:A3alg2}
\end{align}
\end{subequations}
where $i_{2}[A]$, $D_{2}[Q]$, $j_{3}[A]$, and $I_{1,2}[A,Q]$ are
the absolute invariants composed of $A(z)$ and $Q(z)$ as the followings
(cf., Ref.~\cite{Ta03a} and the references cited therein):
\begin{align}
\begin{split}
i_{2}[A]&=12a_{0}a_{4}-3a_{1}a_{3}+a_{2}^{\,2},\qquad
 D_{2}[Q]=4b_{0}b_{2}-b_{1}^{\,2},\\
2j_{3}[A]&=72a_{0}a_{2}a_{4}-27a_{0}a_{3}^{\,2}
 -27a_{1}^{\,2}a_{4}+9a_{1}a_{2}a_{3}-2a_{2}^{\,3},\\
I_{1,2}[A,Q]&=6a_{4}b_{0}^{\,2}-3a_{3}b_{0}b_{1}+2a_{2}b_{0}b_{2}
 +a_{2}b_{1}^{\,2}-3a_{1}b_{1}b_{2}+6a_{0}b_{2}^{\,2}.
\end{split}
\end{align}
One of the most important aspects of the type A $3$-fold SUSY system
(\ref{eqs:A3sys}) is that its components satisfy the third-order
intertwining relations:
\begin{align}
P_{31}^{-}P_{32}^{-}P_{33}^{-}H^{-}=H^{+}P_{31}^{-}P_{32}^{-}
 P_{33}^{-},\quad P_{33}^{+}P_{32}^{+}P_{31}^{+}H^{+}=
 H^{-}P_{33}^{+}P_{32}^{+}P_{31}^{+},
\label{eq:inter}
\end{align}
which are responsible for the first commutation relation in
(\ref{eq:A3alg1}). Before considering the central issues on
the existence of intermediate Hamiltonians, we shall refer to
another transformation property of the system. That is, the type A
$3$-fold SUSY system (\ref{eqs:A3sys}) is transformed under
$W(x)\to-W(x)$ as follows:
\begin{align}
H^{-}\lra H^{+},\quad P_{31}^{-}\lra -P_{33}^{+},\quad
 P_{32}^{-}\lra -P_{32}^{+},\quad P_{33}^{-}\lra -P_{31}^{+},\quad
 P_{3}^{-}\lra -P_{3}^{+}.
\label{eq:Wto-W}
\end{align}
Hence, the system as a whole remains invariant under the transformation
$W(x)\to-W(x)$. In particular, it leaves the pair of intertwining
relations (\ref{eq:inter}) invariant. This transformation however
does not belong to the $GL(2,\bbC)$ transformation since it changes
the sign of $W(x)$ while any $GL(2,\bbC)$ transformation does not
change $W(x)$ at all as shown in (\ref{eq:tfEW}).

For type A $3$-fold SUSY systems, there are essentially three different
patterns in the existence of intermediate Hamiltonians according to
which we shall classify them as follows:\\

\noindent
Class $(1,1)$:
\begin{subequations}
\label{eqs:caseII}
\begin{align}
P_{33}^{-}H^{-}=H^{\rmi1}P_{33}^{-},\qquad
 P_{33}^{+}H^{\rmi1}=H^{-}P_{33}^{+},
\label{eq:caseII1}\\
P_{32}^{-}H^{\rmi1}=H^{\rmj1}P_{32}^{-},\qquad
 P_{32}^{+}H^{\rmj1}=H^{\rmi1}P_{32}^{+},
\label{eq:caseII2}\\
P_{31}^{-}H^{\rmj1}=H^{+}P_{31}^{-},\qquad
 P_{31}^{+}H^{+}=H^{\rmj1}P_{31}^{+}.
\label{eq:caseII3}
\end{align}
\end{subequations}
Class $(0,1)$:
\begin{subequations}
\label{eqs:caseI}
\begin{gather}
P_{32}^{-}P_{33}^{-}H^{-}=H^{\rmj1}P_{32}^{-}P_{33}^{-},
 \qquad P_{33}^{+}P_{32}^{+}H^{\rmj1}=H^{-}P_{33}^{+}P_{32}^{+},
\label{eq:caseI1}\\
P_{31}^{-}H^{\rmj1}=H^{+}P_{31}^{-},\qquad
 P_{31}^{+}H^{+}=H^{\rmj1}P_{31}^{+}.
\label{eq:caseI2}
\end{gather}
\end{subequations}
Class $(1,0)$:
\begin{subequations}
\label{eqs:caseI'}
\begin{gather}
P_{33}^{-}H^{-}=H^{\rmi1}P_{33}^{-},\qquad
 P_{33}^{+}H^{\rmi1}=H^{-}P_{33}^{+},\\
P_{31}^{-}P_{32}^{-}H^{\rmi1}=H^{+}P_{31}^{-}P_{32}^{-},\qquad
 P_{32}^{+}P_{31}^{+}H^{+}=H^{\rmi1}P_{32}^{+}P_{31}^{+}.
\end{gather}
\end{subequations}
That is, in Class $(1,1)$ we have considered a set of intermediate
Hamiltonians $H^{\rmi1}$ and $H^{\rmj1}$ both at the place between
$P_{31}^{\pm}$ and $P_{32}^{\pm}$ and at the place between
$P_{32}^{\pm}$ and $P_{33}^{\pm}$ simultaneously while in Class
$(0,1)$ and Class $(1,0)$ we have considered an intermediate
Hamiltonian at only one intermediate place, at the place between
$P_{31}^{\pm}$ and $P_{32}^{\pm}$ in the former and at the place
between $P_{32}^{\pm}$ and $P_{33}^{\pm}$ in the latter.

But we easily find that the set of the intertwining relations
(\ref{eqs:caseI'}) in Class $(1,0)$ is transformed to the one
(\ref{eqs:caseI}) in Class $(0,1)$ by the transformations
(\ref{eq:Wto-W}) which are caused by $W(x)\to-W(x)$ provided that
$H^{\rmi1}$ is transformed to $H^{\rmj1}$ simultaneously under
the same transformation. We shall hereafter call the set of the
transformations (\ref{eq:Wto-W}) accompanied with the interchange
$H^{\rmj1}\lra H^{\rmi1}$ the \emph{reflective} transformation of a type
A $3$-fold SUSY system. Hence, as long as only one set of
intermediate Hamiltonians is concerned, we only need to consider
Class $(0,1)$ without any loss of generality. When we shall examine
more than one sets of intermediate Hamiltonians, however, we must
consider simultaneously both Class $(0,1)$ and Class $(1,0)$, as we
shall discuss in Sections \ref{sec:exdif} and \ref{sec:cls2}.

\section{Existence of Intermediate Hamiltonians}
\label{sec:exinh}

In this section, we shall derive the necessary and sufficient conditions
for the existence of (at least) one set of intermediate Hamiltonians in
each class classified in the last section. We note that Class $(1,1)$
can be regarded as a special case of either Class $(0,1)$ or Class
$(1,0)$. Thus, we only consider models in Class $(0,1)$ and Class
$(1,0)$ which do not belong to Class $(1,1)$ without any loss of
generality. In the subsequent sections, we shall investigate each
case in details.

\subsection{Conditions for Class $(1,1)$}

The necessary and sufficient condition for satisfying the first formula
(\ref{eq:caseII1}) is that there exists a constant $C_{33}$ such
that $H^{-}$ and $H^{\rmi1}$ are expressed as
\begin{subequations}
\label{eqs:cond5}
\begin{align}
2H^{-}&=P_{33}^{+}P_{33}^{-}+2C_{33}
 =-\del^{2}-W'-E'+W^{2}+2EW+E^{2}+2C_{33},\\
2H^{\rmi1}&=P_{33}^{-}P_{33}^{+}+2C_{33}
 =-\del^{2}+W'+E'+W^{2}+2EW+E^{2}+2C_{33}.
\end{align}
\end{subequations}
Similarly, the necessary and sufficient condition for satisfying
the second formula in (\ref{eq:caseII2}) is that there exists another
constant $C_{32}$ such that $H^{\rmi1}$ and $H^{\rmj1}$ are expressed as
\begin{subequations}
\label{eqs:cond6}
\begin{align}
2H^{\rmi1}&=P_{32}^{+}P_{32}^{-}+2C_{32}
 =-\del^{2}-W'+W^{2}+2C_{32},\\
2H^{\rmj1}&=P_{32}^{-}P_{32}^{+}+2C_{32}
 =-\del^{2}+W'+W^{2}+2C_{32},
\end{align}
\end{subequations}
and the necessary and sufficient condition for satisfying
the third formula in (\ref{eq:caseII3}) is that there exists another
constant $C_{31}$ such that $H^{\rmj1}$ and $H^{+}$ are expressed as
\begin{subequations}
\label{eqs:cond7}
\begin{align}
2H^{\rmj1}&=P_{31}^{+}P_{31}^{-}+2C_{31}
 =-\del^{2}-W'+E'+W^{2}-2EW+E^{2}+2C_{31},\\
2H^{+}&=P_{31}^{-}P_{31}^{+}+2C_{31}
 =-\del^{2}+W'-E'+W^{2}-2EW+E^{2}+2C_{31}.
\end{align}
\end{subequations}
Comparing (\ref{eq:A3Ham}), (\ref{eqs:cond5}), (\ref{eqs:cond6}),
and (\ref{eqs:cond7}) each other, we obtain
\begin{subequations}
\label{eqs:cond8}
\begin{align}
6(W'+EW)+E'+E^{2}&=-6(R+C_{33}),\\
2(W'+EW)+E'+E^{2}&=2(C_{32}-C_{33}),\\
2(W'+EW)-E'-E^{2}&=2(C_{31}-C_{32}),\\
6(W'+EW)-E'-E^{2}&=6(C_{31}+R).
\end{align}
\end{subequations}
Arranging the set of formulas (\ref{eqs:cond8}) and using the relations
(\ref{eq:Q'A''}), we have
\begin{align}
2Q'(z)=C_{33}-C_{31},\quad
 A''(z)=-3(2R+C_{33}+C_{31})=2C_{32}-C_{33}-C_{31}.
\label{eq:cond9}
\end{align}
In terms of the coefficients of polynomials $A(z)$ and $Q(z)$
in (\ref{eqs:polAQ}), the necessary and sufficient conditions
(\ref{eq:cond9}) are written as
\begin{align}
\begin{split}
a_{4}=a_{3}=b_{2}=0,\qquad 2a_{2}=2C_{32}-C_{33}-C_{31},\\
2b_{1}=C_{33}-C_{31},\qquad -3R=C_{33}+C_{32}+C_{31}.
\label{eq:conII}
\end{split}
\end{align}
The last three equalities in (\ref{eq:conII}) just determine the
parameters $C_{31}$, $C_{32}$, and $C_{33}$ in terms of the parameters
of the original type A $3$-fold SUSY system as
\begin{align}
C_{31}=-\frac{a_{2}}{3}-b_{1}-R,\qquad C_{32}=\frac{2a_{2}}{3}-R,
 \qquad C_{33}=-\frac{a_{2}}{3}+b_{1}-R.
\label{eq:C3i}
\end{align}
We note that the first equality in (\ref{eq:conII}) is identical to
the solvability condition of type A $\cN$-fold SUSY systems (cf.,
Ref.~\cite{Ta06a}, Eq.~(6.13)), namely, the necessary and sufficient
condition for the pair of any type A $\cN$-fold SUSY Hamiltonians
$H^{\pm}$ to be not only quasi-solvable but also solvable. Hence,
a type A $3$-fold SUSY system with a set of two intermediate
Hamiltonians $H^{\rmi1}$ and $H^{\rmi2}$ is always solvable, and
conversely a solvable type A $3$-fold SUSY system always admits a set
of two intermediate Hamiltonians.

\subsection{Conditions for Class $(0,1)$}

To consider the first formula in (\ref{eq:caseI1}), we first note that
the second-order operator $P_{32}^{-}P_{33}^{-}$ belongs to a type A
2-fold supercharge
\begin{align}
P_{32}^{-}P_{33}^{-}=\left(\del+W_{1}-\frac{E}{2}\right)
 \left(\del+W_{1}+\frac{E}{2}\right),
\end{align}
with
\begin{align}
W_{1}(x)=W(x)+\frac{E(x)}{2}.
\label{eq:W1}
\end{align}
Hence, the necessary and sufficient conditions for satisfying the first
formula (\ref{eq:caseI1}) are that there exists a second-degree
polynomial $Q_{1}(z)$ in $z$ such that (cf., Ref.~\cite{Ta03a})
\begin{align}
Q_{1}(z)=-z'(x)W_{1}(x),\qquad\frac{\rmd^{3}}{\rmd z^{3}}Q_{1}(z)=0,
\label{eq:Q1}
\end{align}
and that $H^{-}$ and $H^{\rmj1}$ are expressed as
\begin{subequations}
\label{eqs:cond1}
\begin{align}
2H^{-}&=-\frac{\rmd^{2}}{\rmd x^{2}}+W_{1}^{\,2}-\frac{E'}{2}
 +\frac{E^{2}}{4}-2R_{1}-2W_{1}',\\
2H^{\rmj1}&=-\frac{\rmd^{2}}{\rmd x^{2}}+W_{1}^{\,2}-\frac{E'}{2}
 +\frac{E^{2}}{4}-2R_{1}+2W_{1}',
\end{align}
\end{subequations}
where $R_{1}$ is a constant.
The necessary and sufficient condition for satisfying the second formula
(\ref{eq:caseI2}) is that there exists another constant $C_{31}$ such
that $H^{\rmj1}$ and $H^{+}$ are expressed as
\begin{subequations}
\label{eqs:cond2}
\begin{align}
2H^{\rmj1}&=P_{31}^{+}P_{31}^{-}+2C_{31}
 =-\del^{2}-W'+E'+W^{2}-2EW+E^{2}+2C_{31},\\
2H^{+}&=P_{31}^{-}P_{31}^{+}+2C_{31}
 =-\del^{2}+W'-E'+W^{2}-2EW+E^{2}+2C_{31}.
\end{align}
\end{subequations}
Comparing (\ref{eq:A3Ham}), (\ref{eqs:cond1}), and (\ref{eqs:cond2})
each other, and using (\ref{eq:W1}), we obtain
\begin{align}
6(W'+EW)-E'-E^{2}=-12(R-R_{1})=4(C_{31}+R_{1})=6(C_{31}+R),
\end{align}
which is equivalent, in view of the relations (\ref{eq:Q'A''}), to
\begin{align}
-6Q'(z)-A''(z)=-12(R-R_{1})=4(C_{31}+R_{1})=6(C_{31}+R).
\label{eq:cond3}
\end{align}
The condition (\ref{eq:Q1}), if combined with (\ref{eq:defAQ}),
leads to
\begin{align}
A'(z)=2\left[Q(z)-Q_{1}(z)\right],
\label{eq:cond4}
\end{align}
which means that $A'(z)$ is also a polynomial of at most second
degree. In terms of the coefficients of polynomials $A(z)$ and $Q(z)$
in (\ref{eqs:polAQ}), the necessary and sufficient conditions
(\ref{eq:cond3}) and (\ref{eq:cond4}) are written as
\begin{align}
a_{4}=a_{3}+2b_{2}=0,\quad a_{2}+3b_{1}=-2(C_{31}+R_{1}),\quad
 3R=2R_{1}-C_{31}.
\label{eq:condI}
\end{align}
The last two equalities in (\ref{eq:condI}) just determine the constants
$C_{31}$ and $R_{1}$ in terms of the parameters of the original type A
$3$-fold SUSY system as
\begin{align}
C_{31}=-\frac{a_{2}}{3}-b_{1}-R,\qquad
 R_{1}=-\frac{a_{2}}{6}-\frac{b_{1}}{2}+R.
\end{align}
We now find that a model in Class $(0,1)$ which satisfies (\ref{eq:condI})
can also satisfy the condition (\ref{eq:conII}) and thus belong to
Class $(1,1)$ if and only if $a_{3}=b_{2}=0$. Hence, we assume
$a_{3}b_{2}\neq0$ without any loss of generality in the subsequent
analyses of Class $(0,1)$. With the latter assumption, however,
the solvability condition cannot be satisfied inevitably. Hence,
all the models in Class $(0,1)$ are only quasi-solvable but may not
be completely solvable.

Finally, we note that the superHamiltonian and type A $2$-fold
supercharges constructed from $H^{-}$, $H^{\rmj1}$, and
$P_{32}^{-}P_{33}^{-}$ form a type A $2$-fold superalgebra since
$H^{-}$ and $H^{\rmj1}$ are a type A $2$-fold SUSY pair with respect
to the operator $P_{32}^{-}P_{33}^{-}$. In particular, the
anti-commutator of the type A $2$-fold supercharges constructed in
this way reads in components
\begin{align}
P_{33}^{+}P_{32}^{+}P_{32}^{-}P_{33}^{-}=4S_{2}^{(0,1)}(H^{-}+R),
 \qquad P_{32}^{-}P_{33}^{-}P_{33}^{+}P_{32}^{+}=4S_{2}^{(0,1)}(
 H^{\rmj1}+R),
\label{eq:ant01}
\end{align}
where $S_{2}^{(0,1)}$ is a monic polynomial of second-degree given by
\begin{align}
S_{2}^{(0,1)}(t)=t^{2}-\left(\frac{a_{2}}{3}+b_{1}\right)t
 -2(a_{1}-2b_{0})b_{2}-\frac{2}{9}(a_{2}-3b_{1})a_{2}.
\label{eq:S201}
\end{align}

\subsection{Conditions for Class $(1,0)$}

The analysis for Class $(1,0)$ is almost the same as for Class $(0,1)$
in the previous section. That is, the first formula in (\ref{eqs:caseI'})
requires that $H^{-}$ and $H^{\rmi1}$ must form an ordinary SUSY pair
with respect to $P_{33}^{\pm}$:
\begin{subequations}
\begin{align}
2H^{-}&=P_{33}^{+}P_{33}^{-}+2C_{33}
 =-\del^{2}-W'-E'+W^{2}+2EW+E^{2}+2C_{33},\\
2H^{\rmi1}&=P_{33}^{-}P_{33}^{+}+2C_{33}
 =-\del^{2}+W'+E'+W^{2}+2EW+E^{2}+2C_{33},
\end{align}
\end{subequations}
while the second formula in (\ref{eqs:caseI'}) requires that $H^{\rmi1}$
and $H^{+}$ must form a type A $2$-fold SUSY pair with respect to
$P_{31}^{-}P_{32}^{-}$ and its conjugate:
\begin{subequations}
\begin{align}
2H^{\rmi1}&=-\del^{2}+W_{2}^{\,2}-\frac{E'}{2}+\frac{E^{2}}{4}
 -2R_{2}-2W_{2}',\\
2H^{+}&=-\del^{2}+W_{2}^{\,2}-\frac{E'}{2}+\frac{E^{2}}{4}
 -2R_{2}+2W_{2}',
\end{align}
\end{subequations}
where $W_{2}(x)=W(x)-E(x)/2$. By following a similar route as in
the previous section, it is straightforward to show that the above
requirements are satisfied if and only if
\begin{align}
a_{4}=a_{3}-2b_{2}=0,\quad C_{33}=-\frac{a_{2}}{3}+b_{1}-R,\quad
 R_{2}=-\frac{a_{2}}{6}+\frac{b_{1}}{2}+R.
\end{align}
The last two equalities in the above just determine the constants
$C_{33}$ and $R_{2}$ and thus only the first formula stands
essentially as the necessary and sufficient condition for a system
to belong to Class $(1,0)$.
As in the case of Class $(0,1)$, we assume $a_{3}b_{2}\neq0$ without
any loss of generality to prevent a model in Class $(1,0)$ from
belonging also to Class $(1,1)$. Hence, by the assumption, any system
in Class $(1,0)$ is only quasi-solvable but may not be completely
solvable. In addition, the superHamiltonian and type A $2$-fold
supercharges constructed from $H^{\rmi1}$, $H^{+}$, and
$P_{31}^{-}P_{32}^{-}$ form another type A $2$-fold superalgebra
since $H^{\rmi1}$ and $H^{+}$ are a type A $2$-fold SUSY pair
with respect to the operator $P_{31}^{-}P_{32}^{-}$. In
particular, the anti-commutator of the type A $2$-fold
supercharges constructed in this way reads in components
\begin{align}
P_{32}^{+}P_{31}^{+}P_{31}^{-}P_{32}^{-}=4S_{2}^{(1,0)}(H^{\rmi1}+R),
 \qquad P_{31}^{-}P_{32}^{-}P_{32}^{+}P_{31}^{+}=4S_{2}^{(0,1)}(
 H^{+}+R),
\label{eq:ant10}
\end{align}
where $S_{2}^{(1,0)}$ is a monic polynomial of second-degree given by
\begin{align}
S_{2}^{(1,0)}(t)=t^{2}-\left(\frac{a_{2}}{3}-b_{1}\right)t
 +2(a_{1}+2b_{0})b_{2}-\frac{2}{9}(a_{2}+3b_{1})a_{2}.
\label{eq:S210}
\end{align}

\section{Classification: Cases of Only One Set}
\label{sec:cls1}

We are now in a position to classify completely all the possible
type A 3-fold SUSY potentials which admit one and only one set of
intermediate Hamiltonians. In the previous section, we showed
that there are three different classes, Class $(1,1)$, Class $(0,1)$,
and Class $(1,0)$, and that the necessary and sufficient conditions
for the existence of intermediate Hamiltonians are
\begin{align}
&\text{Class $(1,1)$:}\quad a_{4}=a_{3}=b_{2}=0,
\label{eq:c2}\\
&\text{Class $(0,1)$:}\quad a_{4}=a_{3}+2b_{2}=0\quad(a_{3}b_{2}\neq0),
\label{eq:c1}\\
&\text{Class $(1,0)$:}\quad a_{4}=a_{3}-2b_{2}=0\quad(a_{3}b_{2}\neq0),
\label{eq:c1'}
\end{align}
where we have neglected the other irrelevant constraints on the
parameter relations. It is evident that these conditions explicitly
break the $GL(2,\bbC)$ covariance of the original type A $3$-fold
systems. Hence, we cannot apply the same classification scheme as
the one in Ref.~\cite{Ta03a} which employs the full $GL(2,\bbC)$
covariance of the systems. However, we note that there exists
a residual symmetry remained intact in the present cases. In fact,
the transformation formulas (\ref{eq:tfEW}) for the functions $E(x)$
and $W(x)$ tell us that for the set of linear projective transformations
with $\gamma=0$ the functions $E(x)$ and $W(x)$ are both invariant.
The latter fact means in particular that each factor of the type A
3-fold supercharges $P_{3i}^{\pm}$ ($i=1,2,3$) in (\ref{eq:P3i+-})
and thus the intermediate Hamiltonians $H^{\rmi1}$ and $H^{\rmj1}$
introduced through the intertwining relations (\ref{eqs:caseII})--%
(\ref{eqs:caseI'}) as well are all invariant under an arbitrary
$GL(2,\bbC)$ transformation with $\gamma=0$. The set of linear
projective transformations with $\gamma=0$ consists of the set
of (complex) inhomogeneous linear transformations
\begin{align}
z=\alpha w+\beta\quad (\alpha,\beta\in\bbC, \alpha\neq 0),
\label{eq:ilt}
\end{align}
where we have set $\delta=1$ without any loss of generality. For
an arbitrary inhomogeneous linear transformation (\ref{eq:ilt}),
the transformation matrices of the parameters $a_{i}$ ($i=0,\dots,4$)
and $b_{i}$ ($i=0,1,2$) given by (\ref{eq:tfas}) and (\ref{eq:tfbs})
respectively become triangle, and all of the conditions
(\ref{eq:c2})--(\ref{eq:c1'}) are covariant, that is,
\begin{align}
a_{4}=a_{3}=b_{2}=0&\quad\Rightarrow\quad
 \hat{a}_{4}=\hat{a}_{3}=\hat{b}_{2}=0.\\
a_{4}=a_{3}\pm2b_{2}=0&\quad\Rightarrow\quad\hat{a}_{4}=0,
 \ \hat{a}_{3}\pm2\hat{b}_{2}=\alpha(a_{3}\pm2b_{2})=0.
\end{align}
Therefore, all type A 3-fold SUSY systems with one set of intermediate
Hamiltonians are classified by considering the equivalence class under
the set of inhomogeneous linear transformations (\ref{eq:ilt}).
We now easily show that the representatives of $A(z)$ under this
equivalence class in each case can be chosen, by noting the constraint
$a_{3}\neq0$ in Class $(0,1)$ and Class $(1,0)$, as listed in
Table~\ref{tb:cclas}.
\begin{table}
\begin{center}
\[
\tabcolsep=10pt
\begin{tabular}{lll}
\hline
 Case & Class $(1,1)$ & Class $(0,1)$, $(1,0)$\\
\hline
 I & $a/2$ & \\
 II & $2z$ & \\
 II' & & $2z^{3}$\\
 III & $z^{2}/2$ & \\
 IV & $2a(z^{2}-1)$ & \\
 IV' & & $2az^{2}(z+1)$\\
 V & & $2z^{3}-g_{2}z/2-g_{3}/2$\\
\hline
\end{tabular}
\]
\caption{The complex classification scheme of type A 3-fold SUSY
 models with one set of intermediate Hamiltonians. In the above,
$g_{2},g_{3}\in\bbC$ satisfy $g_{2}^{\,3}\neq27 g_{3}^{\,2}$ and
$a\in\bbC$ is an arbitrary constant.}
\label{tb:cclas}
\end{center}
\end{table}
We note that Case II and Case II' (Case IV and Case IV', respectively)
are transformed to each other by a projective transformation of
$GL(2,\bbC)$ but not by any inhomogeneous linear transformation
(\ref{eq:ilt}). That is the reason why we must treat them as separate
cases though they are classified as one equivalent case in the
classification of general type A $\cN$-fold SUSY models~\cite{Ta03a}.
The still arbitrarily determinable constant $a$ in Table~\ref{tb:cclas}
can be fixed by considering the scaling relations under the scale
transformations of the parameter space spanned by $\{a_{i},b_{i},R\}$:
\begin{subequations}
\label{eqs:scale}
\begin{align}
z(x;\nu a_{i},\nu b_{i},\nu R)&=z(\rnu x;a_{i},b_{i},R),\\
E(x;\nu a_{i},\nu b_{i},\nu R)&=\rnu E(\rnu x;a_{i},b_{i},R),\\
W(x;\nu a_{i},\nu b_{i},\nu R)&=\rnu W(\rnu x;a_{i},b_{i},R),\\
V(x;\nu a_{i},\nu b_{i},\nu R)&=\nu V(\rnu x;a_{i},b_{i},R).
\end{align}
\end{subequations}
Hence, in what follows we set $a=1$ in all the cases without any
loss of generality. For the classification, we recall the fact that
Class $(0,1)$ and Class $(1,0)$ are connected with each other by
the reflective transformation. Hence, we shall present only models
belonging to Class $(0,1)$. We shall show in each case the change
of variable $z=z(x)$ and the two functions $E(x)$ and $W(x)$
determined by (\ref{eq:defAQ}) as well as the potential parts of
the $3$-fold SUSY pair Hamiltonians $H^{\pm}$ and of the intermediate
Hamiltonian(s) $H^{\rmi1}$ and/or $H^{\rmj1}$ determined by
(\ref{eqs:cond5})--(\ref{eqs:cond7}) in Class $(1,1)$ and by
(\ref{eqs:cond1})--(\ref{eqs:cond2}) in Class $(0,1)$.

\subsection{Classification of Class $(1,1)$}
\label{ssec:cls11}

\noindent
I) $A(z)=1/2$.\vspace{5pt}\\
\textit{Functions}:
\begin{align}
z(x)=x,\qquad E(x)=0,\qquad W(x)=-b_{1}x-b_{0}.
\end{align}
\textit{Potentials}:
\begin{subequations}
\label{eqs:111}
\begin{align}
2V^{-}(x)&=(b_{1}x+b_{0})^{2}+3b_{1}-2R,\\
2V^{\rmi1}(x)&=(b_{1}x+b_{0})^{2}+b_{1}-2R,\\
2V^{\rmj1}(x)&=(b_{1}x+b_{0})^{2}-b_{1}-2R,\\
2V^{+}(x)&=(b_{1}x+b_{0})^{2}-3b_{1}-2R.
\end{align}
\end{subequations}
In this case, all the potentials (\ref{eqs:111}) are harmonic
oscillators.\\
\newpage
\noindent
II) $A(z)=2z$.\vspace{5pt}\\
\textit{Functions}:
\begin{align}
z(x)=x^{2},\qquad E(x)=\frac{1}{x},\qquad
 W(x)=-\frac{b_{1}}{2}x-\frac{b_{0}}{2x}.
\end{align}
\textit{Potentials}:
\begin{subequations}
\label{eqs:112}
\begin{align}
V^{-}(x)&=\frac{b_{1}^{\,2}}{8}x^{2}+\frac{(b_{0}-2)(b_{0}-4)}{8x^{2}}
 +\frac{b_{1}(b_{0}+3)}{4}-R,\\
V^{\rmi1}(x)&=\frac{b_{1}^{\,2}}{8}x^{2}+\frac{b_{0}(b_{0}-2)}{8x^{2}}
 +\frac{b_{1}(b_{0}+1)}{4}-R,\\
V^{\rmj1}(x)&=\frac{b_{1}^{\,2}}{8}x^{2}+\frac{b_{0}(b_{0}+2)}{8x^{2}}
 +\frac{b_{1}(b_{0}-1)}{4}-R,\\
V^{+}(x)&=\frac{b_{1}^{\,2}}{8}x^{2}+\frac{(b_{0}+2)(b_{0}+4)}{8x^{2}}
 +\frac{b_{1}(b_{0}-3)}{4}-R.
\end{align}
\end{subequations}
In this case, all the potentials (\ref{eqs:112}) are radial harmonic
oscillators with centrifugal potentials.\\

\noindent
III) $A(z)=z^{2}/2$.\vspace{5pt}\\
\textit{Functions}:
\begin{align}
z(x)=\rme^{x},\qquad E(x)=1,\qquad W(x)=-b_{1}-b_{0}\rme^{-x}.
\end{align}
\textit{Potentials}:
\begin{subequations}
\label{eqs:113}
\begin{align}
2V^{-}(x)&=b_{0}(2b_{1}-3)\rme^{-x}+b_{0}^{\,2}\rme^{-2x}-2\bar{R},\\
2V^{\rmi1}(x)&=b_{0}(2b_{1}-1)\rme^{-x}+b_{0}^{\,2}\rme^{-2x}
 -2\bar{R},\\
2V^{\rmj1}(x)&=b_{0}(2b_{1}+1)\rme^{-x}+b_{0}^{\,2}\rme^{-2x}
 -2\bar{R},\\
2V^{+}(x)&=b_{0}(2b_{1}+3)\rme^{-x}+b_{0}^{\,2}\rme^{-2x}-2\bar{R},
\end{align}
\end{subequations}
where $\bar{R}=R-b_{1}^{\,2}/2-1/3$ is a constant.
In this case, all the potentials (\ref{eqs:113}) are Morse potentials.\\

\noindent
IV) $A(z)=2(z^{2}-1)$.\vspace{5pt}\\
\textit{Functions}:
\begin{align}
z(x)=\cosh 2x,\qquad E(x)=\frac{2\cosh 2x}{\sinh 2x},\qquad
 W(x)=-\frac{b_{1}\cosh 2x+b_{0}}{2\sinh 2x}.
\end{align}
\textit{Potentials}:
\begin{subequations}
\label{eqs:114}
\begin{align}
V^{-}(x)&=\frac{2b_{0}(b_{1}-6)\cosh 2x+b_{0}^{\,2}+(b_{1}-4)(b_{1}-8)
 }{8\sinh^{2}2x}-\bar{R},\\
V^{\rmi1}(x)&=\frac{2b_{0}(b_{1}-2)\cosh 2x+b_{0}^{\,2}+b_{1}(b_{1}-4)
 }{8\sinh^{2}2x}-\bar{R},\\
V^{\rmj1}(x)&=\frac{2b_{0}(b_{1}+2)\cosh 2x+b_{0}^{\,2}+b_{1}(b_{1}+4)
 }{8\sinh^{2}2x}-\bar{R},\\
V^{+}(x)&=\frac{2b_{0}(b_{1}+6)\cosh 2x+b_{0}^{\,2}+(b_{1}+4)(b_{1}+8)
 }{8\sinh^{2}2x}-\bar{R},
\end{align}
\end{subequations}
where $\bar{R}=R-b_{1}^{\,2}/8-4/3$ is a constant.
In this case, all the potentials (\ref{eqs:114}) are P\"{o}schl--Teller
potentials. We note that a system of Scarf potentials can be obtained
by the choice $A(z)=2(z^{2}+1)$ which is connected with the representative
$A(z)=2(z^{2}-1)$ by the combination of a complex linear transformation
(\ref{eq:ilt}) and a scale transformation (\ref{eqs:scale}).
%

It is interesting to observe that all the potentials given by
(\ref{eqs:111}), (\ref{eqs:112}), (\ref{eqs:113}), and (\ref{eqs:114})
are shape invariant as can be easily
checked by some obvious scaling of parameters~\cite{CKS95}.
It bears mention that the schemes of SUSY quantum mechanics
\cite{Wi81,CKS95,Ju96,Ba00} are consistent with the factorization
method~\cite{Sc40a}, intertwining relationships~\cite{Da1882},
and the shape invariance condition~\cite{Ge83}. The latter was
first utilized by Gendenshtein~\cite{Ge83} to track down solvable
potentials. To be shape invariant the partner potentials while
sharing a similar coordinate dependence can at most differ in
the presence of some parameters as precisely has happened in
the potentials above.

\subsection{Classification of Class $(0,1)$}
\label{ssec:cls10}

\noindent
II') $A(z)=2z^{3}$.\vspace{5pt}\\
\textit{Functions}:
\begin{align}
z(x)=\frac{1}{x^{2}},\qquad E(x)=-\frac{3}{x},\qquad
 W(x)=-\frac{1}{2x}+\frac{b_{1}}{2}x+\frac{b_{0}}{2}x^{3}.
\end{align}
\textit{Potentials}:
\begin{subequations}
\label{eqs:102}
\begin{align}
V^{-}(x)&=\frac{b_{0}^{\,2}}{8}x^{6}+\frac{b_{1}b_{0}}{4}x^{4}
 +\frac{b_{1}^{\,2}-20b_{0}}{8}x^{2}+\frac{3}{8x^{2}}-b_{1}-R,\\
V^{\rmj1}(x)&=\frac{b_{0}^{\,2}}{8}x^{6}+\frac{b_{1}b_{0}}{4}x^{4}
 +\frac{b_{1}^{\,2}+8b_{0}}{8}x^{2}+\frac{35}{8x^{2}}-R,\\
V^{+}(x)&=\frac{b_{0}^{\,2}}{8}x^{6}+\frac{b_{1}b_{0}}{4}x^{4}
 +\frac{b_{1}^{\,2}+16b_{0}}{8}x^{2}+\frac{15}{8x^{2}}
 +\frac{b_{1}}{2}-R.
\end{align}
\end{subequations}
In this case, all the potentials (\ref{eqs:102}) are well-known
quasi-solvable sextic anharmonic oscillators~\cite{TU87}.\\

\noindent
IV') $A(z)=2z^{2}(z+1)$.\vspace{5pt}\\
\textit{Functions}:
\begin{align}
z(x)=\frac{1}{\sinh^{2}x},\quad E(x)=-\frac{3+2\sinh^{2}x}{
 \sinh x\cosh x},\quad W(x)=\frac{b_{0}\sinh^{4}x
 +b_{1}\sinh^{2}x-1}{2\sinh x\cosh x}.
\end{align}
\textit{Potentials}:
\begin{subequations}
\label{eqs:104}
\begin{align}
V^{-}(x)=&\:\frac{b_{0}^{\,2}}{8}\cosh^{4}x+\frac{(2b_{1}-3b_{0}-12)
 b_{0}}{8}\cosh^{2}x\notag\\
&-\frac{(b_{1}-b_{0}+3)(b_{1}-b_{0}+5)}{8\cosh^{2}x}
 +\frac{3}{8\sinh^{2}x}+\frac{b_{0}}{2}-\bar{R},\\
V^{\rmj1}(x)=&\:\frac{b_{0}^{\,2}}{8}\cosh^{4}x+\frac{(2b_{1}-3b_{0}
 +4)b_{0}}{8}\cosh^{2}x\notag\\
&-\frac{(b_{1}-b_{0}-1)(b_{1}-b_{0}+1)}{8\cosh^{2}x}
 +\frac{35}{8\sinh^{2}x}-\frac{b_{0}}{2}-\bar{R},\\
V^{+}(x)=&\:\frac{b_{0}^{\,2}}{8}\cosh^{4}x+\frac{(2b_{1}-3b_{0}
 +12)b_{0}}{8}\cosh^{2}x\notag\\
&-\frac{(b_{1}-b_{0}-3)(b_{1}-b_{0}-1)}{8\cosh^{2}x}
 +\frac{15}{8\sinh^{2}x}-b_{0}-\bar{R},
\end{align}
\end{subequations}
where $\bar{R}=R-(b_{1}-b_{0})(b_{1}-3b_{0})/8-4/3$ is a constant.
In this case, all the potentials (\ref{eqs:104}) are quasi-solvable
deformed P\"{o}schl--Teller or Scarf potentials~\cite{Za90}.\\

\noindent
V) $A(z)=2z^{3}-g_{2}z/2-g_{3}/2$.\vspace{5pt}\\
\textit{Functions}:
\begin{align}
z(x)=\wp(x),\qquad E(x)=\frac{\wp''(x)}{\wp'(x)},\qquad
 W(x)=\frac{\wp(x)^{2}-b_{1}\wp(x)-b_{0}}{\wp'(x)}.
\end{align}
\textit{Potentials}:
\begin{subequations}
\label{eqs:105}
\begin{align}
V^{-}(x)&=\frac{b_{1}\wp(x)+\bar{b}_{0}}{2\wp'(x)^{2}}
 \left[b_{1}\wp(x)+\bar{b}_{0}-\frac{10}{3}\wp''(x)\right]
 -\frac{8}{9}\wp(x)+\frac{91}{18}\wp(2x)+\frac{3b_{1}}{2}-R,\\
V^{\rmj1}(x)&=\frac{b_{1}\wp(x)+\bar{b}_{0}}{2\wp'(x)^{2}}
 \left[b_{1}\wp(x)+\bar{b}_{0}+\frac{2}{3}\wp''(x)\right]
 +\frac{40}{9}\wp(x)-\frac{5}{18}\wp(2x)-\frac{b_{1}}{2}-R,\\
V^{+}(x)&=\frac{b_{1}\wp(x)+\bar{b}_{0}}{2\wp'(x)^{2}}
 \left[b_{1}\wp(x)+\bar{b}_{0}+\frac{8}{3}\wp''(x)\right]
 +\frac{10}{9}\wp(x)+\frac{55}{18}\wp(2x)-\frac{3b_{1}}{2}-R.
\end{align}
\end{subequations}
In the above, $\wp(x)$ is the Weierstrass elliptic function and
$\bar{b}_{0}=b_{0}-g_{2}/12$ is another parameter. The first term
of each the potential is a rational function of $\wp(x)$ by the
formulas $\wp'(x)^{2}=4\wp(x)^{3}-g_{2}\wp(x)-g_{3}$ and
$\wp''(x)=6\wp(x)^{2}-g_{2}/2$. In this case, all the potentials
(\ref{eqs:105}) are quasi-solvable one-body elliptic $BC$ type
Inozemtsev potentials (cf., Ref.~\cite{Ta04} and the references
cited therein).

\section{Different Sets of Intermediate Hamiltonians}
\label{sec:exdif}

In this section, we shall study under what conditions a type A 3-fold
SUSY system can have more than one sets of intermediate Hamiltonians.
The latter possibility originates from the fact that each factor of
a type A $\cN$-fold supercharge in a factorized form is not invariant
under a subset of the $GL(2,\bbC)$ transformations although any type A
$\cN$-fold supercharge as a whole is invariant under all the
$GL(2,\bbC)$ transformations~\cite{Ta03a}. For the $\cN=3$ case,
we easily check the latter fact directly from the definition
(\ref{eq:P3i+-}) and the transformation formulas (\ref{eq:tfEW}):
\begin{align}
P_{31}^{\pm}[\hW,\hE]=P_{31}^{\pm}[W,E]+\frac{2\gamma z'}{\gamma z
 -\alpha},\quad P_{33}^{\pm}[\hW,\hE]=P_{33}^{\pm}[W,E]
 -\frac{2\gamma z'}{\gamma z-\alpha}.
\end{align}
Hence, the factors $P_{31}^{\pm}$ and $P_{33}^{\pm}$ in $P_{3}^{\pm}$
are in fact not invariant under any $GL(2,\bbC)$ transformation
with $\gamma\neq0$. It means in particular that intermediate
Hamiltonians $H^{\rmi1}$ and/or $H^{\rmj1}$ defined by the relations
(\ref{eqs:caseII}), (\ref{eqs:caseI}), or (\ref{eqs:caseI'}) could
not be invariant under any transformation with $\gamma\neq0$, even
if they exist after the transformation. In fact, the existence of
intermediate Hamiltonians after a transformation is not guaranteed
automatically since the conditions for the existence (\ref{eq:c2})
or (\ref{eq:c1}) are not preserved under any $GL(2,\bbC)$
transformation with $\gamma\neq0$ as was already discussed in
Section~\ref{sec:cls1}. Therefore, another different set of
intermediate Hamiltonians exists if there is a $GL(2,\bbC)$
transformation with $\gamma\neq0$ for which the transformed
parameters $\hat{a}_{i}$ and $\hat{b}_{i}$ given by (\ref{eq:tfas})
and (\ref{eq:tfbs}) also satisfy the existence conditions
(\ref{eq:c2}) or (\ref{eq:c1}). Furthermore, if there exist
simultaneously such $n(>1)$ $GL(2,\bbC)$ transformations characterized
by the sets $\{\alpha_{i},\beta_{i},\gamma_{i},\delta_{i}\}_{i=1}^{n}$
satisfying $\gamma_{i}\neq0$ and $\alpha_{i}/\gamma_{i}\neq\alpha_{j}/
\gamma_{j}$ for all $i\neq j$, then the corresponding $n$ sets of
intermediate Hamiltonians are different from each other since the
formula (\ref{eq:tfEW}) tells us that the deformation is characterized
by the one parameter $\alpha/\gamma$. We shall hereafter say that
two $GL(2,\bbC)$ transformations with $\gamma_{i},\gamma_{j}\neq0$ are
\emph{inequivalent} if $\alpha_{i}/\gamma_{i}\neq\alpha_{j}/\gamma_{j}$.

In the classification of the systems which admit one set of
intermediate Hamiltonians in Section~\ref{sec:cls1}, we considered
the two different classes, namely, Class $(1,1)$ where an intermediate
Hamiltonian exists at each of the two intermediate positions in the
factorized type A $3$-fold supercharge and Class $(0,1)$ where an
intermediate Hamiltonian exists only at one intermediate position.
And any system belonging to Class $(1,0)$ can be obtained from the
corresponding system belonging to Class $(0,1)$ by the reflective
transformation $W(x)\to-W(x)$.
Accordingly, we can consider a type A $3$-fold SUSY system which
admits simultaneously $m$ different sets of intermediate Hamiltonians
$\{H^{\rmi k},H^{\rmj k}\}_{k=1}^{m}$ of Class $(1,1)$ and $n$
different sets of an intermediate Hamiltonian $\{H^{\rmj k}\}_{k=m+1
}^{m+n}$ of Class $(0,1)$. We shall call such a class of systems Class
$(m,m+n)$ with an obvious implication of the terminology. It is evident
that any system which admits simultaneously $m$ different sets of
intermediate Hamiltonians $\{H^{\rmi k},H^{\rmj k}\}_{k=1}^{m}$ of
Class $(1,1)$ and $n$ different sets of an intermediate Hamiltonian
$\{H^{\rmi k}\}_{k=m+1}^{m+n}$ of Class $(1,0)$, which would be called
to belong Class $(m+n,m)$, can be obtained from the corresponding
system belonging to Class $(m,m+n)$ by the reflective transformation
(with the obvious accompaniment of the interchanges $H^{\rmj k}\lra
H^{\rmi k}$ for $k>1$).
To investigate each class systematically, we first note that any system
which belongs to Class $(m,m+n)$ with $m,n>0$ always belongs to Class
$(1,1)$ since the conditions for Class $(1,1)$ is stricter than for
Class $(0,1)$. In other words, an arbitrary system in Class $(m,m+n)$
with $m>0$ is a special case of one of the four systems in Class
$(1,1)$ classified in Section~\ref{ssec:cls11} and thus is always
solvable in particular. Hence, only Class $(0,n)$ with $n>1$, which is
an abbreviation for Class $(0,0+n)$, can have a quasi-solvable system
which must be a special case of one of the three systems in Class
$(0,1)$ classified in Section~\ref{ssec:cls10}.

In addition to these classes, there could exist \emph{hybrid} classes.
That is, there is the possibility that a type A $3$-fold SUSY system
admits simultaneously different sets of intermediate Hamiltonians some
of which belong to Class $(0,1)$ and the others of which belong to
Class $(1,0)$. These systems, if exist, could belong to neither Class
$(m,m+n)$ nor Class $(0,n)$ since Class $(0,1)$ and Class $(1,0)$ are,
as already mentioned, connected by the reflective transformation but not by
any $GL(2,\bbC)$ transformation. We can thus consider a class of
systems belonging to Class $(m,m+n)$ which admit simultaneously $l$
additional different sets of intermediate Hamiltonians
$\{H^{\rmi k}\}_{k=m+1}^{m+l}$ of Class $(1,0)$. We shall call such a
class of systems Class $(m+l,m+n)$. We can assume $n\geq l$ without
any loss of generality since the reflective transformation interchanges
Class $(m+n,m+l)$ and Class $(m+l,m+n)$. By following an argument
similar to in the last paragraph, we shall separate the hybrid classes
in two, the one is Class $(m+l,m+n)$ with $m,l,n>0$ which is a special
case of Class $(m,m+n)$ with $m,n>0$ and the other is Class $(l;n)$,
which is an abbreviation for Class $(0+l,0+n)$, with $n\geq l\geq1$
and is a special case of Class $(0,n)$ with $n\geq1$.

In the subsequent sections, we shall study Class $(m,m+n)$ with
$m,n>0$ and Class $(0,n)$ with $n\geq1$ separately.

\subsection{Conditions for Class $(m,m+n)$ with $m,n>0$}
\label{ssec:conmnm}

The necessary and sufficient conditions for the existence of a set of
intermediate Hamiltonians of Class $(1,1)$ are given by (\ref{eq:c2}).
A system in Class $(1,1)$ also belongs to Class $(m,m+n)$ if there
exist simultaneously $n$ inequivalent $GL(2,\bbC)$ transformations
for which the conditions (\ref{eq:c1}) for Class $(0,1)$ are satisfied
by the transformed coefficients $\hat{a}_{i}$ ($i=0,\dots,4$) and
$\hat{b}_{i}$ ($i=0,1,2$), that is,
\begin{align}
\hat{a}_{4}=\hat{a}_{3}+2\hat{b}_{2}=0\quad
 (\hat{a}_{3}\hat{b}_{2}\neq0),
\label{eq:con10}
\end{align}
and in addition if there exist simultaneously $m-1$ additional
inequivalent $GL(2,\bbC)$ transformations for which the conditions
(\ref{eq:c2}) for Class $(1,1)$ are satisfied by $\hat{a}_{i}$ and
$\hat{b}_{i}$, that is,
\begin{align}
\hat{a}_{4}=\hat{a}_{3}=\hat{b}_{2}=0.
\label{eq:con11}
\end{align}
{}From the transformation formulas (\ref{eq:tfas}) and
(\ref{eq:tfbs}), we see that when the condition (\ref{eq:c2})
is satisfied, the transformed coefficients $\hat{a}_{4}$,
$\hat{a}_{3}$, and $\hat{b}_{2}$ respectively read
\begin{subequations}
\label{eqs:tfab1}
\begin{align}
\Delta^{2}\hat{a}_{4}&=\alpha^{2}\gamma^{2}a_{2}
 +\alpha\gamma^{3}a_{1}+\gamma^{4}a_{0},\\
\Delta^{2}\hat{a}_{3}&=2\alpha\gamma(\alpha\delta+\beta\gamma)a_{2}
 +\gamma^{2}(3\alpha\delta+\beta\gamma)a_{1}+4\gamma^{3}\delta a_{0},\\
\Delta\hat{b}_{2}&=\alpha\gamma b_{1}+\gamma^{2}b_{0}.
\end{align}
\end{subequations}
Hence, except for the trivial case $\gamma=0$, the conditions
(\ref{eq:con10}) are satisfied if and only if
\begin{align}
\alpha^{2}a_{2}+\alpha\gamma a_{1}+\gamma^{2}a_{0}=0,\qquad
 2\alpha a_{2}+\gamma a_{1}-2\alpha b_{1}-2\gamma b_{0}=0,
\label{eq:con12}
\end{align}
with $\hat{a}_{3}\hat{b}_{2}\neq0$, and the conditions
(\ref{eq:con11}) are satisfied if and only if
\begin{align}
\alpha^{2}a_{2}+\alpha\gamma a_{1}+\gamma^{2}a_{0}=0,\qquad
 2\alpha a_{2}+\gamma a_{1}=0,\qquad
 \alpha b_{1}+\gamma b_{0}=0.
\label{eq:con13}
\end{align}
In addition, a system of Class $(m,m+n)$ also belongs to Class
$(m+l,m+n)$ if there exist simultaneously additional $l$ inequivalent
$GL(2,\bbC)$ transformations for which the transformed coefficients
$\hat{a}_{i}$ and $\hat{b}_{i}$ satisfy the conditions (\ref{eq:c1'})
for Class $(1,0)$, namely,
\begin{align}
\hat{a}_{4}=\hat{a}_{3}-2\hat{b}_{2}=0\quad(\hat{a}_{3}\hat{b}_{2}
 \neq0).
\label{eq:con14}
\end{align}
By the transformation formulas (\ref{eqs:tfab1}), they are satisfied
if and only if (with $\hat{a}_{3}\hat{b}_{2}\neq0$)
\begin{align}
\alpha^{2}a_{2}+\alpha\gamma a_{1}+\gamma^{2}a_{0}=0,\qquad
 2\alpha a_{2}+\gamma a_{1}+2\alpha b_{1}+2\gamma b_{0}=0.
\label{eq:con15}
\end{align}
In Section~\ref{ssec:clsmnm}, we shall investigate the conditions
(\ref{eq:con12}), (\ref{eq:con13}), and (\ref{eq:con15}) in each
case separately.

\subsection{Conditions for Class $(0,n)$ with $n>1$}
\label{ssec:conn0}

The necessary and sufficient conditions for the existence of an
intermediate Hamiltonian of Class $(0,1)$ are given by (\ref{eq:c1}).
A system in Class $(0,1)$ also belongs to Class $(0,n)$ if there
exist simultaneously $n-1$ inequivalent $GL(2,\bbC)$ transformations
for which the transformed coefficients $\hat{a}_{i}$ and $\hat{b}_{i}$
satisfy the condition for Class $(0,1)$, namely, Eq.~(\ref{eq:con10}).
{}From the transformation formulas (\ref{eq:tfas}) and (\ref{eq:tfbs}),
we see that when the condition (\ref{eq:c1}) is satisfied, the
transformed coefficients $\hat{a}_{4}$, $\hat{a}_{3}$, and
$\hat{b}_{2}$ respectively read
\begin{subequations}
\label{eqs:tfab2}
\begin{align}
\Delta^{2}\hat{a}_{4}&=\alpha^{3}\gamma a_{3}+\alpha^{2}\gamma^{2}
 a_{2}+\alpha\gamma^{3}a_{1}+\gamma^{4}a_{0},\\
\Delta^{2}\hat{a}_{3}&=\alpha^{2}(\alpha\delta+3\beta\gamma)a_{3}
 +2\alpha\gamma(\alpha\delta+\beta\gamma)a_{2}+\gamma^{2}(3\alpha
 \delta+\beta\gamma)a_{1}+4\gamma^{3}\delta a_{0},\\
\Delta\hat{b}_{2}&=-\alpha^{2}a_{3}/2+\alpha\gamma b_{1}
 +\gamma^{2}b_{0}.
\end{align}
\end{subequations}
Hence, we obtain
\begin{align}
\Delta^{2}(\hat{a}_{3}+2\hat{b}_{2})=&\:4\alpha^{2}\beta\gamma
 a_{3}+2\alpha\gamma(\alpha\delta+\beta\gamma)a_{2}+\gamma^{2}
 (3\alpha\delta+\beta\gamma)a_{1}+4\gamma^{3}\delta a_{0}\notag\\
&+2\Delta\alpha\gamma b_{1}+2\Delta\gamma^{2}b_{0}.
\end{align}
Therefore, the condition (\ref{eq:con10}) is satisfied, except for
the trivial case $\gamma=0$, if and only if
\begin{subequations}
\label{eqs:con16}
\begin{align}
\alpha^{3}a_{3}+\alpha^{2}\gamma a_{2}+\alpha\gamma^{2}a_{1}
 +\gamma^{3}a_{0}=0,\\
4\alpha^{2}a_{3}+2\alpha\gamma a_{2}+\gamma^{2}a_{1}
 -2\alpha\gamma b_{1}-2\gamma^{2}b_{0}=0,
\end{align}
\end{subequations}
with $\hat{a}_{3}\hat{b}_{2}\neq0$. The second condition is derived
by the elimination of $a_{0}$. On the other hand, a system in Class
$(0,n)$ also belongs to Class $(l;n)$ if there exist simultaneously
$l$ additional inequivalent $GL(2,\bbC)$ transformations for which
the transformed coefficients $\hat{a}_{i}$ and $\hat{b}_{i}$ satisfy
the conditions for Class $(1,0)$, namely, Eq.~(\ref{eq:con14}).
By the transformation formulas (\ref{eqs:tfab2}) they are satisfied
if and only if (with $\hat{a}_{3}\hat{b}_{2}\gamma\neq0$)
\begin{subequations}
\label{eqs:con17}
\begin{align}
\alpha^{3}a_{3}+\alpha^{2}\gamma a_{2}+\alpha\gamma^{2}a_{1}
 +\gamma^{3}a_{0}=0,\\
2\alpha^{2}a_{3}+2\alpha\gamma a_{2}+\gamma^{2}a_{1}
 +2\alpha\gamma b_{1}+2\gamma^{2}b_{0}=0.
\end{align}
\end{subequations}
In Section~\ref{ssec:clsn0}, we shall investigate the conditions
(\ref{eqs:con16}) and (\ref{eqs:con17}) in each case separately.

\section{Classification: Cases of More Than One Sets}
\label{sec:cls2}

Now that we have derived the existence conditions for another different
set of intermediate Hamiltonians in a general form, we are in a
position to proceed a detailed analysis for each case classified in
Section~\ref{sec:cls1}. In what follows, we first investigate the
systems which belong to Class $(m,m+n)$ with $m,n>0$ and next the ones
which belong to Class $(0,n)$ with $n>1$. All the former models are
not only quasi-solvable but also solvable since they are special cases
of Class $(1,1)$. On the other hand, all the latter models are merely
quasi-solvable since we have excluded either cases of $a_{3}=b_{2}=0$
or of $\hat{a}_{3}=\hat{b}_{2}=0$ in Class $(0,n)$.

\subsection{Classification of Class $(m,m+n)$ with $m,n>0$}
\label{ssec:clsmnm}

\noindent
I) $A(z)=1/2$:\vspace{5pt}\\
In this case, both the conditions (\ref{eq:con12}) and (\ref{eq:con13})
only have a trivial solution $\gamma=0$. Thus, the system admits no
different sets of intermediate Hamiltonians.\\

\noindent
II) $A(z)=2z$:\vspace{5pt}\\
In this case, the conditions (\ref{eq:con13}) only have a trivial
solution but the conditions (\ref{eq:con12}) have one set of
non-trivial solutions
\begin{align}
\alpha=0,\quad b_{0}=1\quad\text{with}\quad\hat{a}_{3}=-2\hat{b}_{2}=
 2\gamma/\beta(\neq0).
\label{eq:sol2}
\end{align}
Thus, the system admits another set of an intermediate Hamiltonian of
Class $(0,1)$ and belongs to Class $(1,2)$, which is an abbreviation
for Class $(1,1+1)$. On the other hand, the conditions (\ref{eq:con15})
also have one set of non-trivial solutions
\begin{align}
\alpha=0,\quad b_{0}=-1\quad\text{with}\quad\hat{a}_{3}=2\hat{b}_{2}=
 2\gamma/\beta(\neq0).
\end{align}
However, the latter $GL(2,\bbC)$ transformation $\alpha=0$ is equivalent
to the one in (\ref{eq:sol2}), and the latter solution $b_{0}=-1$ is
not compatible with $b_{0}=1$ in (\ref{eq:sol2}). Hence, the system
does not admit a realization of Class $(1+1,1+1)$.\\

\noindent
\textit{Functions}:
\begin{align}
z(x)=x^{2},\quad E(x)=\frac{1}{x},\quad\hE(x)=-\frac{3}{x},\quad
 W(x)=\hW(x)=-\frac{b_{1}}{2}x-\frac{1}{2x}.
\label{eq:EW212}
\end{align}
\textit{Potentials}:
\begin{subequations}
\label{eqs:212}
\begin{align}
V^{-}(x)&=\frac{b_{1}^{\,2}}{8}x^{2}+\frac{3}{8x^{2}}+b_{1}-R,\\
V^{\rmi1}(x)&=\frac{b_{1}^{\,2}}{8}x^{2}-\frac{1}{8x^{2}}
 +\frac{b_{1}}{2}-R,\qquad
V^{\rmj1}(x)=\frac{b_{1}^{\,2}}{8}x^{2}+\frac{3}{8x^{2}}-R,\\
V^{\rmj2}(x)&=\frac{b_{1}^{\,2}}{8}x^{2}+\frac{35}{8x^{2}}-R,
\label{eq:Vj22}\\
V^{+}(x)&=\frac{b_{1}^{\,2}}{8}x^{2}+\frac{15}{8x^{2}}
 -\frac{b_{1}}{2}-R.
\end{align}
\end{subequations}
It is interesting to note that the formulas (\ref{eqs:cond5}) and
(\ref{eqs:cond6}) are not valid with $\hE(x)$ and $\hW(x)$ given in
(\ref{eq:EW212}); the potential term $V^{\rmi2}(x)$, for instance,
calculated by (\ref{eqs:cond5}) and calculated by (\ref{eqs:cond6})
do not coincide with each other. Hence, the system admits only
$H^{\rmj2}$ but not $H^{\rmi2}$ as the second set of intermediate
Hamiltonians. All the potentials (\ref{eqs:212}) including the newly
appeared intermediate one $V^{\rmj2}(x)$ in (\ref{eq:Vj22})
belong to the class of radial harmonic oscillators with a
\emph{particular} angular momentum.\\

\noindent
III) $A(z)=z^{2}/2$:\vspace{5pt}\\
In this case, only the conditions (\ref{eq:con13}) have one set of
non-trivial solutions
\begin{align}
\alpha=0,\quad b_{0}=0\quad\text{with}\quad\hat{a}_{3}=\hat{b}_{2}=0.
\label{eq:con223}
\end{align}
Thus, the system admits another set of intermediate Hamiltonians of
Class $(1,1)$ and belongs to Class $(2,2)$.
However, when the conditions (\ref{eq:con223}) are satisfied, then
$E(x)=-\hE(x)=1$ and $W(x)=\hW(x)=-b_{1}$ so that all the potentials
are just an identical constant. Hence, this case is nothing more than
a trivial system.\\

\noindent
IV) $A(z)=2(z^{2}-1)$:\vspace{5pt}\\
In this case, the conditions (\ref{eq:con13}) only have a trivial
solution but the conditions (\ref{eq:con12}) have two sets of
non-trivial solutions
\begin{align}
\alpha=\pm\gamma,\quad b_{0}=\mp b_{1}\pm2\quad\text{with}\quad
 \hat{a}_{3}=-2\hat{b}_{2}=-4\gamma/(\delta\mp\beta)(\neq0).
\label{eq:sol4}
\end{align}
Thus, the system admits another set of intermediate Hamiltonians of
Class $(0,1)$ corresponding to each of the solutions and  belongs to
Class $(1,2)$. In addition, the system with the specific values of
parameters $b_{1}=2$ and $b_{0}=0$ admits the two different solutions
\emph{simultaneously} and thus can have additional two different sets
of intermediate Hamiltonians of Class $(0,1)$ corresponding to the two
solutions. In the latter case, the system belongs to Class $(1,3)$.
On the other hand, the conditions (\ref{eq:con15}) also have two sets
of non-trivial solutions
\begin{align}
\alpha=\pm\gamma,\quad b_{0}=\mp b_{1}\mp2\quad\text{with}\quad
 \hat{a}_{3}=2\hat{b}_{2}=-4\gamma/(\delta\mp\beta)(\neq0).
\end{align}
They are compatible with (\ref{eq:sol4}) if and only if $b_{1}=0$ and
$b_{0}\pm2$ with the inequivalent transformations, namely,
$\alpha=\pm\gamma$ for the former and $\alpha=\mp\gamma$ for the latter.
In these particular two cases, the system belongs to Class $(1+1,1+1)$.\\

\noindent
IV-1) Class $(1,2)$\vspace{5pt}\\
\textit{Parameters}:
\begin{align}
b_{0}=\mp(b_{1}-2),\quad\hat{a}_{3}=-2\hat{b}_{2}=-\frac{4\gamma}{
 \delta\mp\beta},\quad\hat{a}_{2}=\frac{2(\beta\pm5\delta)}{\beta\mp
 \delta},\quad\hat{b}_{1}=-b_{1}+\frac{4\delta}{\delta\mp\beta}.
\end{align}
\textit{Functions}:
\begin{align}
\begin{split}
z(x)=\cosh 2x,\qquad W(x)=\hW(x)=-\frac{b_{1}\cosh 2x+b_{0}}{2\sinh 2x},\\
E(x)=\frac{2\cosh 2x}{\sinh 2x},\qquad\hE(x)=\frac{2\cosh 2x}{\sinh 2x}
 -\frac{4\sinh 2x}{\cosh 2x\mp1}.
\label{eq:EW214}
\end{split}
\end{align}
\textit{Potentials}:
\begin{subequations}
\label{eqs:214}
\begin{align}
V^{-}(x)&=\frac{\mp(b_{1}-2)(b_{1}-6)\cosh 2x+b_{1}^{\,2}-8b_{1}+18}{
 4\sinh^{2}2x}-\bar{R},\\
V^{\rmi1}(x)&=\frac{\mp(b_{1}-2)^{2}\cosh 2x+b_{1}^{\,2}-4b_{1}+2}{
 4\sinh^{2}2x}-\bar{R},\\
V^{\rmj1}(x)&=\frac{\mp(b_{1}^{\,2}-4)\cosh 2x+b_{1}^{\,2}+2}{
 4\sinh^{2}2x}-\bar{R},\\
V^{\rmj2}(x)&=\frac{\mp(b_{1}^{\,2}-4)\cosh 2x+b_{1}^{\,2}+2}{
 4\sinh^{2}2x}\pm\frac{8}{\cosh 2x\mp1}-\bar{R},
\label{eq:Vj24}\\
V^{+}(x)&=\frac{\mp(b_{1}-2)(b_{1}+6)\cosh 2x+b_{1}^{\,2}+4b_{1}+18}{
 4\sinh^{2}2x}-\bar{R},
\end{align}
\end{subequations}
where $\bar{R}=R-b_{1}^{\,2}/8-4/3$ is a constant. The formulas
(\ref{eqs:cond5}) and (\ref{eqs:cond6}) are again not valid with
$\hE(x)$ and $\hW(x)$ given in (\ref{eq:EW214}) and thus the system
admits only $H^{\rmj2}$ but not $H^{\rmi2}$ as the second set of
intermediate Hamiltonians. All the potentials (\ref{eqs:214}) but the
newly appeared intermediate one $V^{\rmj2}(x)$ in (\ref{eq:Vj24})
belong to the class of P\"{o}schl--Teller potential with only one free
parameter. The deformation term in (\ref{eq:Vj24}) is reminiscent of
the generalized P\"{o}schl--Teller potentials constructed in
Ref.~\cite{BQR09}. It is worth noting that the deformed potential
(\ref{eq:Vj24}) is connected with the other shape-invariant potentials
by the intertwining relations (\ref{eqs:caseI}) and thus is almost
isospectral to them and is in particular solvable.\\

\noindent
IV-2) Class $(1,3)$\vspace{5pt}\\
\textit{Parameters}:
\begin{align}
\begin{split}
&b_{1}=2,\quad\hat{a}_{3}=-2\hat{b}_{2}=-\frac{4\gamma}{\delta-\beta},
 \quad\hat{a}_{2}=\frac{2(\beta+5\delta)}{\beta-\delta},\quad
 \hat{b}_{1}=\frac{2(\delta+\beta)}{\delta-\beta},\\
&b_{0}=0,\quad\Hat{\Hat{a}}_{3}=-2\Hat{\Hat{b}}_{2}=-\frac{4\gamma}{
 \delta+\beta},
 \quad\Hat{\Hat{a}}_{2}=\frac{2(\beta-5\delta)}{\beta+\delta},
 \quad\Hat{\Hat{b}}_{1}=\frac{2(\delta-\beta)}{\delta+\beta}.
\end{split}
\end{align}
\textit{Functions}:
\begin{align}
\begin{split}
&z(x)=\cosh 2x,\quad E(x)=\frac{2\cosh 2x}{\sinh 2x},\quad
 W(x)=\hW(x)=\Hat{\hW}(x)=-\frac{\cosh 2x}{\sinh 2x},\\
&\hE(x)=\frac{2\cosh 2x}{\sinh 2x}-\frac{4\sinh 2x}{\cosh 2x-1},\qquad
 \Hat{\hE}(x)=\frac{2\cosh 2x}{\sinh 2x}-\frac{4\sinh 2x}{\cosh 2x+1}.
\end{split}
\end{align}
\textit{Potentials}:
\begin{subequations}
\label{eqs:314}
\begin{align}
V^{-}(x)&=\frac{3}{2\sinh^{2}2x}-\bar{R},\\
V^{\rmi1}(x)&=-\frac{1}{2\sinh^{2}2x}-\bar{R},\qquad
 V^{\rmj1}(x)=\frac{3}{2\sinh^{2}2x}-\bar{R},\\
V^{\rmj2}(x)&=\frac{3}{2\sinh^{2}2x}+\frac{8}{\cosh 2x-1}-\bar{R},\\
V^{\rmj3}(x)&=\frac{3}{2\sinh^{2}2x}-\frac{8}{\cosh 2x+1}-\bar{R},\\
V^{+}(x)&=\frac{15}{2\sinh^{2}2x}-\bar{R},
\end{align}
\end{subequations}
where $\bar{R}=R-11/6$ is a constant. There are essentially no free
parameters in this case.\\

\noindent
IV-3) Class $(1+1,1+1)$\vspace{5pt}\\
\textit{Parameters}:
\begin{align}
\begin{split}
b_{1}=0,\quad\hat{a}_{3}=-2\hat{b}_{2}=-\frac{4\gamma}{\delta\mp\beta},
 \quad\hat{a}_{2}=\frac{2(\beta\pm5\delta)}{\beta\mp\delta},\quad
 \hat{b}_{1}=\frac{4\delta}{\delta\mp\beta},\\
b_{0}=\pm2,\quad\Hat{\Hat{a}}_{3}=2\Hat{\Hat{b}}_{2}=-\frac{4\gamma}{
 \delta\pm\beta},
 \quad\Hat{\Hat{a}}_{2}=\frac{2(\beta\mp5\delta)}{\beta\pm\delta},
 \quad\Hat{\Hat{b}}_{1}=\frac{4\delta}{\delta\pm\beta}.
\end{split}
\end{align}
\textit{Functions}:
\begin{align}
\begin{split}
&z(x)=\cosh 2x,\quad E(x)=\frac{2\cosh 2x}{\sinh 2x},\quad
 W(x)=\hW(x)=\Hat{\hW}(x)=\mp\frac{1}{\sinh 2x},\\
&\hE(x)=\frac{2\cosh 2x}{\sinh 2x}-\frac{4\sinh 2x}{\cosh 2x\mp1},\qquad
 \Hat{\hE}(x)=\frac{2\cosh 2x}{\sinh 2x}-\frac{4\sinh 2x}{\cosh 2x\pm1}.
\end{split}
\end{align}
\textit{Potentials}:
\begin{subequations}
\label{eqs:224}
\begin{align}
V^{-}(x)&=\frac{\mp6\cosh 2x+9}{2\sinh^{2}2x}-\bar{R},\\
V^{\rmi1}(x)&=-\frac{\mp2\cosh 2x+1}{2\sinh^{2}2x}-\bar{R},\qquad
 V^{\rmj1}(x)=\frac{\pm2\cosh 2x+1}{2\sinh^{2}2x}-\bar{R},\\
V^{\rmi2}(x)&=\frac{\mp2\cosh 2x+1}{2\sinh^{2}2x}
 \mp\frac{8}{\cosh 2x\pm1}-\bar{R},\\
V^{\rmj2}(x)&=\frac{\pm2\cosh 2x+1}{2\sinh^{2}2x}
 \pm\frac{8}{\cosh 2x\mp1}-\bar{R},\\
V^{+}(x)&=\frac{\pm6\cosh 2x+9}{2\sinh^{2}2x}-\bar{R},
\end{align}
\end{subequations}
where $\bar{R}=R-4/3$ is a constant. As in the previous case, there
are essentially no free parameters in this case.

\subsection{Classification of Class $(0,n)$ with $n>1$}
\label{ssec:clsn0}

\noindent
II') $A(z)=2z^{3}$:\vspace{5pt}\\
In this case, the conditions (\ref{eqs:con16}) have one set of
non-trivial solutions $\alpha=b_{0}=0$ but with $\hat{a}_{3}=
\hat{b}_{2}=0$ which should be excluded. In fact, the system
corresponding to the latter solutions is identical with the one
in the case II, which belongs to Class $(1,2)$, already found in
the previous section, Eqs.~(\ref{eqs:212}).\\

\noindent
IV') $A(z)=2z^{2}(z+1)$:\vspace{5pt}\\
In this case, the conditions (\ref{eqs:con16}) have two sets of
non-trivial solutions, the one is
\begin{align}
\alpha=-\gamma,\quad b_{1}=b_{0}-2\quad\text{with}\quad
 \hat{a}_{3}=-2\hat{b}_{2}=2\gamma/(\beta+\delta)(\neq0),
\label{eq:sol4'}
\end{align}
but the other is
\begin{align}
\alpha=b_{0}=0\quad\text{with}\quad\hat{a}_{3}=\hat{b}_{2}=0,
\label{eq:sol4'2}
\end{align}
and thus should be discarded. Indeed, the choice of the latter
solution simply leads to the system of Class $(1,2)$, and together
with the former solution (\ref{eq:sol4'}), to the system of Class
$(1,3)$ which are identical with the systems (\ref{eqs:214}) and
(\ref{eqs:314}) respectively in the case IV already found in
the previous section. Hence, only the solution (\ref{eq:sol4'})
leads to a new system which belongs to Class $(0,2)$.

On the other hand, the conditions (\ref{eqs:con17}) also have
two sets of non-trivial solutions, the one is identical with
(\ref{eq:sol4'2}) to be discarded while the other is
\begin{align}
\alpha=-\gamma,\quad b_{1}=b_{0}\quad\text{with}\quad
 \hat{a}_{3}=2\hat{b}_{2}=2\gamma/(\beta+\delta)(\neq0).
\label{eq:sol4'3}
\end{align}
Hence, the Class $(0,1)$ system (\ref{eqs:104}) admits an
intermediate Hamiltonian of Class $(1,0)$ and thus belongs
to Class $(1;1)$. However, the $GL(2,\bbC)$ transformation
$\alpha=-\gamma$ of the latter solution is equivalent to
the one in (\ref{eq:sol4'}). Hence, the system does not admit
a realization of Class $(1;2)$. We note that the choice of
the two solutions (\ref{eq:sol4'2}) and (\ref{eq:sol4'3})
leads to the system of Class $(1+1,1+1)$ which is identical
with the system (\ref{eqs:224}).\\

\noindent
IV'-1) Class $(0,2)$\vspace{5pt}\\
\textit{Parameters}:
\begin{align}
b_{1}=b_{0}-2,\quad\hat{a}_{3}=-2\hat{b}_{2}=\frac{2\gamma}{
 \beta+\delta},\quad\hat{a}_{2}=\frac{2(\delta-2\beta)}{\delta
 +\beta},\quad\hat{b}_{1}=-b_{0}-\frac{2\delta}{\beta+\delta}.
\end{align}
\textit{Functions}:
\begin{align}
\begin{split}
&z(x)=\frac{1}{\sinh^{2}x},\quad W(x)=\hW(x)=\frac{b_{0}\sinh^{2}x
 \cosh^{2}x-2\sinh^{2}x-1}{2\sinh x\cosh x},\\
&E(x)=-\frac{3+2\sinh^{2}x}{\sinh x\cosh x},\qquad
 \hE(x)=\frac{1-2\sinh^{2}x}{\sinh x\cosh x}.
\end{split}
\end{align}
\textit{Potentials}:
\begin{subequations}
\begin{align}
V^{-}(x)=&\:\frac{b_{0}^{\,2}}{8}\cosh^{4}x-\frac{(b_{0}+16)b_{0}
 }{8}\cosh^{2}x-\frac{3}{8\cosh^{2}x}+\frac{3}{8\sinh^{2}x}
 +\frac{11}{6}+b_{0}-R,\\
V^{\rmj1}(x)=&\:\frac{b_{0}^{\,2}}{8}\cosh^{4}x-\frac{b_{0}^{\,2}
 }{8}\cosh^{2}x-\frac{3}{8\cosh^{2}x}+\frac{35}{8\sinh^{2}x}
 +\frac{11}{6}-R,\\
V^{\rmj2}(x)=&\:\frac{b_{0}^{\,2}}{8}\cosh^{4}x-\frac{b_{0}^{\,2}
 }{8}\cosh^{2}x-\frac{35}{8\cosh^{2}x}+\frac{3}{8\sinh^{2}x}
 +\frac{11}{6}-R,\\
V^{+}(x)=&\:\frac{b_{0}^{\,2}}{8}\cosh^{4}x-\frac{(b_{0}-8)b_{0}
 }{8}\cosh^{2}x-\frac{15}{8\cosh^{2}x}+\frac{15}{8\sinh^{2}x}
 +\frac{11}{6}-\frac{b_{0}}{2}-R.
\end{align}
\end{subequations}
IV'-2) Class $(1;1)$\vspace{5pt}\\
\textit{Parameters}:
\begin{align}
b_{1}=b_{0},\quad\hat{a}_{3}=2\hat{b}_{2}=\frac{2\gamma}{\beta+\delta},
 \quad\hat{a}_{2}=\frac{2(\delta-2\beta)}{\delta+\beta},\quad
 \hat{b}_{1}=-b_{0}-\frac{2\beta}{\beta+\delta}.
\end{align}
\textit{Functions}:
\begin{align}
\begin{split}
&z(x)=\frac{1}{\sinh^{2}x},\quad W(x)=\hW(x)=\frac{b_{0}\sinh^{2}x
 \cosh^{2}x-1}{2\sinh x\cosh x},\\
&E(x)=-\frac{3+2\sinh^{2}x}{\sinh x\cosh x},\qquad
 \hE(x)=\frac{1-2\sinh^{2}x}{\sinh x\cosh x}.
\end{split}
\end{align}
\textit{Potentials}:
\begin{subequations}
\begin{align}
V^{-}(x)=&\:\frac{b_{0}^{\,2}}{8}\cosh^{4}x-\frac{(b_{0}+12)b_{0}
 }{8}\cosh^{2}x-\frac{15}{8\cosh^{2}x}+\frac{3}{8\sinh^{2}x}
 +\frac{4}{3}+\frac{b_{0}}{2}-R,\\
V^{\rmj1}(x)=&\:\frac{b_{0}^{\,2}}{8}\cosh^{4}x-\frac{(b_{0}-4)
 b_{0}}{8}\cosh^{2}x+\frac{1}{8\cosh^{2}x}+\frac{35}{8\sinh^{2}x}
 +\frac{4}{3}-\frac{b_{0}}{2}-R,\\
V^{\rmi1}(x)=&\:\frac{b_{0}^{\,2}}{8}\cosh^{4}x-\frac{(b_{0}+4)
 b_{0}}{8}\cosh^{2}x-\frac{35}{8\cosh^{2}x}-\frac{1}{8\sinh^{2}x}
 +\frac{4}{3}-R,\\
V^{+}(x)=&\:\frac{b_{0}^{\,2}}{8}\cosh^{4}x-\frac{(b_{0}-12)b_{0}
 }{8}\cosh^{2}x-\frac{3}{8\cosh^{2}x}+\frac{15}{8\sinh^{2}x}
 +\frac{4}{3}-b_{0}-R.
\end{align}
\end{subequations}

\noindent
V) $A(z)=2z^{3}-g_{2}z/2-g_{3}/2$:\vspace{5pt}\\
In this case, the conditions (\ref{eqs:con16}) have three sets of
non-trivial solutions
\begin{align}
\alpha=e_{i}\gamma,\quad b_{0}=-e_{i}b_{1}+4e_{i}^{\,2}-\frac{g_{2}}{4}
 \quad\text{with}\quad\hat{a}_{3}=-2\hat{b}_{2}=\frac{\wp''(\omega_{i})
 \gamma}{\beta-e_{i}\delta}(\neq0),
\label{eq:sol51}
\end{align}
where each $e_{i}=\wp(\omega_{i})$ ($i=1,2,3$) is the value of
the Weierstrass elliptic function at the half of the fundamental period
$2\omega_{i}$ which satisfies the third-degree algebraic equation
\begin{align}
4e_{i}^{\,3}-g_{2}e_{i}-g_{3}=0.
\end{align}
Thus, the system admits another set of intermediate Hamiltonians of
Class $(0,1)$ corresponding to each of the solutions and in those cases
the system belongs to Class $(0,2)$. In addition, the system with
the specific values of parameters $b_{1}=4(e_{i}+e_{j})$ and
$b_{0}=-(e_{i}+e_{j})^{2}-3e_{i}e_{j}$ ($i\neq j$) admits
the two different solutions \emph{simultaneously} and thus can have
additional two different sets of intermediate Hamiltonians of Class
$(0,1)$ corresponding to the two solutions. In the latter case, the
system belongs to Class $(0,3)$. We note, however, that the three
different solutions are incompatible simultaneously and hence any
Class $(0,n)$ with $n>3$ cannot be realized.
On the other hand, the conditions (\ref{eqs:con17}) also have three
sets of non-trivial solutions
\begin{align}
\alpha=e_{i}\gamma,\quad b_{0}=-e_{i}b_{1}-2e_{i}^{\,2}
 +\frac{g_{2}}{4}\quad\text{with}\quad\hat{a}_{3}=2\hat{b}_{2}
 =\frac{\wp''(\omega_{i})\gamma}{\beta-e_{i}\delta}(\neq0).
\label{eq:sol52}
\end{align}
Hence, the Class $(0,1)$ system (\ref{eqs:105}) also admits an
intermediate Hamiltonian of Class $(1,0)$ and thus belongs to Class
$(1;1)$. In addition, a choice of one solution $\alpha=e_{i}\gamma$
in (\ref{eq:sol51}) and another $\alpha=e_{j}\gamma$ ($j\neq i$) in
(\ref{eq:sol52}) is compatible with $b_{1}=2e_{i}$ and $b_{0}=
e_{i}^{\,2}-e_{i}e_{j}-e_{j}^{\,2}$. In the latter case, the system
possesses two additional intermediate Hamiltonians the one belongs to
Class $(0,1)$ and the other to Class $(1,0)$, and thus it is a member
of Class $(1,2)$. However, a choice of three solutions, e.g., two
$\alpha=e_{i}\gamma, e_{j}\gamma$ in (\ref{eq:sol51}) and one
$\alpha=e_{k}\gamma$ in (\ref{eq:sol52}) with $i\neq j\neq k\neq i$,
conflict with the assumption of non-degeneracy
$g_{2}^{\,3}\neq27g_{3}^{\,2}$ (for the latter example,
$e_{i}=-2e_{j}$ must hold). Therefore, the hybrid classes such
as Class $(1,3)$ and Class $(2;2)$ cannot be realized anymore.\\

\noindent
V-1) Class $(0,2)$\vspace{5pt}\\
\textit{Parameters}:
\begin{align}
\begin{split}
&b_{0}=-e_{i}b_{1}+4e_{i}^{\,2}-\frac{g_{2}}{4},
 \qquad\hat{b}_{1}=-b_{1}+\frac{4e_{i}^{\,2}\beta-(12e_{i}^{\,3}
 +g_{3})\delta}{2e_{i}(\beta-e_{i}\delta)},\\
&\hat{a}_{3}=-2\hat{b}_{2}=\frac{\wp''(\omega_{i})
 \gamma}{\beta-e_{i}\delta},\qquad
\hat{a}_{2}=3\frac{4e_{i}^{\,2}\beta^{2}+g_{3}\beta\delta-(4e_{i}^{\,3}
 +g_{3})e_{i}\delta^{2}}{2e_{i}(\beta-e_{i}\delta)^{2}}.
\end{split}
\label{eq:p205}
\end{align}
\textit{Functions}:
\begin{align}
\begin{split}
&z(x)=\wp(x),\quad W(x)=\hW(x)=\frac{\wp(x)^{2}-b_{1}
 \wp(x)-b_{0}}{\wp'(x)},\\
&E(x)=\frac{\wp''(x)}{\wp'(x)},\qquad \hE(x)=\frac{\wp''(x)
 }{\wp'(x)}-\frac{2\wp'(x)}{\wp(x)-e_{i}}.\\
\end{split}
\end{align}
\textit{Potentials}:
\begin{subequations}
\label{eqs:205}
\begin{align}
V^{-}(x)=&\:\frac{b_{1}\wp(x)+\bar{b}_{0}}{2\wp'(x)^{2}}
 \left[b_{1}\wp(x)+\bar{b}_{0}-\frac{10}{3}\wp''(x)\right]
 -\frac{8}{9}\wp(x)+\frac{91}{18}\wp(2x)+\frac{3b_{1}}{2}-R,\\
V^{\rmj1}(x)=&\:\frac{b_{1}\wp(x)+\bar{b}_{0}}{2\wp'(x)^{2}}
 \left[b_{1}\wp(x)+\bar{b}_{0}+\frac{2}{3}\wp''(x)\right]
 +\frac{40}{9}\wp(x)-\frac{5}{18}\wp(2x)-\frac{b_{1}}{2}-R,\\
V^{\rmj2}(x)=&\:\frac{b_{1}\wp(x)+\bar{b}_{0}}{2\wp'(x)^{2}}
 \left[b_{1}\wp(x)+\bar{b}_{0}+\frac{2}{3}\wp''(x)\right]
 +\frac{4}{9}\wp(x)-\frac{5}{18}\wp(2x)\notag\\
&+\frac{2\wp''(\omega_{i})}{\wp(x)-e_{i}}+4e_{i}-\frac{b_{1}}{2}
 -R,\\
V^{+}(x)=&\:\frac{b_{1}\wp(x)+\bar{b}_{0}}{2\wp'(x)^{2}}
 \left[b_{1}\wp(x)+\bar{b}_{0}+\frac{8}{3}\wp''(x)\right]
 +\frac{10}{9}\wp(x)+\frac{55}{18}\wp(2x)-\frac{3b_{1}}{2}-R,
\end{align}
\end{subequations}
where $\bar{b}_{0}=b_{0}-g_{2}/12=-e_{i}b_{1}+2\wp''(\omega_{i})/3$.
The first term of each the potential can be expressed solely in terms
of $\wp(x)$. With the latter value of $\bar{b}_{0}$, we have
\begin{align}
\lefteqn{
\frac{b_{1}\wp(x)+\bar{b}_{0}}{2\wp'(x)^{2}}\left[b_{1}\wp(x)
 +\bar{b}_{0}+C\wp''(x)\right]}\hspace{30pt}\notag\\
&=\frac{[3b_{1}(\wp(x)-e_{i})+2\wp''(\omega_{i})]^{2}}{72
 \prod_{l=1}^{3}(\wp(x)-e_{l})}+\frac{C}{12}\sum_{l=1}^{3}
 \frac{3b_{1}(\wp(x)-e_{i})+2\wp''(\omega_{i})}{\wp(x)-e_{l}},
\end{align}
where and hereafter $i\neq j\neq k\neq i$ ($i,j,k=1,2,3$).
\vspace{8pt}\\

\noindent
V-2) Class $(0,3)$\vspace{5pt}\\
\textit{Parameters}: $\hat{a}_{i}$ and $\hat{b}_{i}$ are the same
 as (\ref{eq:p205}).
\begin{align}
\begin{split}
&b_{1}=-4e_{k},\quad b_{0}=-e_{k}^{\,2}-3e_{i}e_{j},
 \quad\Hat{\Hat{b}}_{1}=-b_{1}+\frac{4e_{j}^{\,2}\beta-(12e_{j}^{\,3}
 +g_{3})\delta}{2e_{j}(\beta-e_{j}\delta)},\\
&\Hat{\Hat{a}}_{3}=-2\Hat{\Hat{b}}_{2}=\frac{\wp''(\omega_{j})
 \gamma}{\beta-e_{j}\delta},\qquad
\Hat{\Hat{a}}_{2}=3\frac{4e_{j}^{\,2}\beta^{2}+g_{3}\beta\delta
 -(4e_{j}^{\,3}+g_{3})e_{j}\delta^{2}}{2e_{j}(\beta-e_{j}\delta)^{2}}.
\end{split}
\label{eq:p305}
\end{align}
\textit{Functions}:
\begin{align}
\begin{split}
&z(x)=\wp(x),\qquad W(x)=\hW(x)=\Hat{\hW}(x)=\frac{\wp(x)^{2}+4e_{k}
 \wp(x)+e_{k}^{\,2}-3e_{i}e_{j}}{\wp'(x)},\\
&E(x)=\frac{\wp''(x)}{\wp'(x)},\quad
 \hE(x)=\frac{\wp''(x)}{\wp'(x)}-\frac{2\wp'(x)}{\wp(x)-e_{i}},
 \quad\Hat{\hE}(x)=\frac{\wp''(x)}{\wp'(x)}-\frac{2\wp'(x)
 }{\wp(x)-e_{j}}.
\end{split}
\end{align}
\textit{Potentials}:
\begin{subequations}
\label{eqs:305}
\begin{align}
V^{-}(x)=&\:\frac{b_{1}\wp(x)+\bar{b}_{0}}{2\wp'(x)^{2}}
 \left[b_{1}\wp(x)+\bar{b}_{0}-\frac{10}{3}\wp''(x)\right]
 -\frac{8}{9}\wp(x)+\frac{91}{18}\wp(2x)+\frac{3b_{1}}{2}-R,\\
V^{\rmj1}(x)=&\:\frac{b_{1}\wp(x)+\bar{b}_{0}}{2\wp'(x)^{2}}
 \left[b_{1}\wp(x)+\bar{b}_{0}+\frac{2}{3}\wp''(x)\right]
 +\frac{40}{9}\wp(x)-\frac{5}{18}\wp(2x)-\frac{b_{1}}{2}-R,\\
V^{\rmj2}(x)=&\:\frac{b_{1}\wp(x)+\bar{b}_{0}}{2\wp'(x)^{2}}
 \left[b_{1}\wp(x)+\bar{b}_{0}+\frac{2}{3}\wp''(x)\right]
 +\frac{4}{9}\wp(x)-\frac{5}{18}\wp(2x)\notag\\
&+\frac{2\wp''(\omega_{i})}{\wp(x)-e_{i}}+4e_{i}-\frac{b_{1}}{2}
 -R,\\
V^{\rmj3}(x)=&\:\frac{b_{1}\wp(x)+\bar{b}_{0}}{2\wp'(x)^{2}}
 \left[b_{1}\wp(x)+\bar{b}_{0}+\frac{2}{3}\wp''(x)\right]
 +\frac{4}{9}\wp(x)-\frac{5}{18}\wp(2x)\notag\\
&+\frac{2\wp''(\omega_{j})}{\wp(x)-e_{j}}+4e_{j}-\frac{b_{1}}{2}
 -R,\\
V^{+}(x)=&\:\frac{b_{1}\wp(x)+\bar{b}_{0}}{2\wp'(x)^{2}}
 \left[b_{1}\wp(x)+\bar{b}_{0}+\frac{8}{3}\wp''(x)\right]
 +\frac{10}{9}\wp(x)+\frac{55}{18}\wp(2x)-\frac{3b_{1}}{2}-R.
\end{align}
\end{subequations}
With the values of $b_{1}$ and $b_{0}$ in (\ref{eq:p305}), the first
term of each the potential can be expressed solely in terms of
$\wp(x)$ as
\begin{align}
\lefteqn{
\frac{b_{1}\wp(x)+\bar{b}_{0}}{2\wp'(x)^{2}}\left[b_{1}\wp(x)
 +\bar{b}_{0}+C\wp''(x)\right]}\hspace{30pt}\notag\\
&=\frac{2[3e_{k}\wp(x)+e_{k}^{\,2}+2e_{i}e_{j}]^{2}}{9
 \prod_{l=1}^{3}(\wp(x)-e_{l})}-\frac{C}{3}\sum_{l=1}^{3}
 \frac{3e_{k}\wp(x)+e_{k}^{\,2}+2e_{i}e_{j}}{\wp(x)-e_{l}}.
\end{align}

\noindent
V-3) Class $(1;1)$\vspace{5pt}\\
\textit{Parameters}:
\begin{align}
\begin{split}
&b_{0}=-e_{i}b_{1}-2e_{i}^{\,2}+\frac{g_{2}}{4},
 \qquad\hat{b}_{1}=-b_{1}+\frac{4e_{i}^{\,2}\beta+(4e_{i}^{\,3}
 +g_{3})\delta}{2e_{i}(\beta-e_{i}\delta)},\\
&\hat{a}_{3}=2\hat{b}_{2}=\frac{\wp''(\omega_{i})
 \gamma}{\beta-e_{i}\delta},\qquad
\hat{a}_{2}=3\frac{4e_{i}^{\,2}\beta^{2}+g_{3}\beta\delta-(4e_{i}^{\,3}
 +g_{3})e_{i}\delta^{2}}{2e_{i}(\beta-e_{i}\delta)^{2}}.
\end{split}
\end{align}
\textit{Functions}:
\begin{align}
\begin{split}
&z(x)=\wp(x),\quad W(x)=\hW(x)=\frac{\wp(x)^{2}-b_{1}
 \wp(x)-b_{0}}{\wp'(x)},\\
&E(x)=\frac{\wp''(x)}{\wp'(x)},\qquad \hE(x)=\frac{\wp''(x)
 }{\wp'(x)}-\frac{2\wp'(x)}{\wp(x)-e_{i}},\\
\end{split}
\end{align}
\textit{Potentials}:
\begin{subequations}
\begin{align}
V^{-}(x)=&\:\frac{b_{1}\wp(x)+\bar{b}_{0}}{2\wp'(x)^{2}}
 \left[b_{1}\wp(x)+\bar{b}_{0}-\frac{10}{3}\wp''(x)\right]
 -\frac{8}{9}\wp(x)+\frac{91}{18}\wp(2x)+\frac{3b_{1}}{2}-R,\\
V^{\rmi1}(x)=&\:\frac{b_{1}\wp(x)+\bar{b}_{0}}{2\wp'(x)^{2}}
 \left[b_{1}\wp(x)+\bar{b}_{0}-\frac{4}{3}\wp''(x)\right]
 -\frac{2}{9}\wp(x)+\frac{7}{18}\wp(2x)\notag\\
&+\frac{2\wp''(\omega_{i})}{\wp(x)-e_{i}}+4e_{i}+\frac{b_{1}}{2}
 -R,\\
V^{\rmj1}(x)=&\:\frac{b_{1}\wp(x)+\bar{b}_{0}}{2\wp'(x)^{2}}
 \left[b_{1}\wp(x)+\bar{b}_{0}+\frac{2}{3}\wp''(x)\right]
 +\frac{40}{9}\wp(x)-\frac{5}{18}\wp(2x)-\frac{b_{1}}{2}-R,\\
V^{+}(x)=&\:\frac{b_{1}\wp(x)+\bar{b}_{0}}{2\wp'(x)^{2}}
 \left[b_{1}\wp(x)+\bar{b}_{0}+\frac{8}{3}\wp''(x)\right]
 +\frac{10}{9}\wp(x)+\frac{55}{18}\wp(2x)-\frac{3b_{1}}{2}-R,
\end{align}
\end{subequations}
where $\bar{b}_{0}=b_{0}-g_{2}/12=-e_{i}b_{1}-\wp''(\omega_{i})/3$.
The first term of each the potential can be expressed solely in terms
of $\wp(x)$. With the latter value of $\bar{b}_{0}$, we have
\begin{align}
\lefteqn{
\frac{b_{1}\wp(x)+\bar{b}_{0}}{2\wp'(x)^{2}}\left[b_{1}\wp(x)
 +\bar{b}_{0}+C\wp''(x)\right]}\hspace{30pt}\notag\\
&=\frac{[3b_{1}(\wp(x)-e_{i})-\wp''(\omega_{i})]^{2}}{72
 \prod_{l=1}^{3}(\wp(x)-e_{l})}+\frac{C}{12}\sum_{l=1}^{3}
 \frac{3b_{1}(\wp(x)-e_{i})-\wp''(\omega_{i})}{\wp(x)-e_{l}}.
\end{align}

\noindent
V-4) Class $(1;2)$\vspace{5pt}\\
\textit{Parameters}: $\hat{a}_{i}$ and $\hat{b}_{i}$ are the same
 as (\ref{eq:p205}).
\begin{align}
\begin{split}
&b_{1}=2e_{i},\quad b_{0}=e_{i}^{\,2}+e_{j}e_{k},
 \quad\Hat{\Hat{b}}_{1}=-b_{1}+\frac{4e_{j}^{\,2}\beta+(4e_{j}^{\,3}
 +g_{3})\delta}{2e_{j}(\beta-e_{j}\delta)},\\
&\Hat{\Hat{a}}_{3}=2\Hat{\Hat{b}}_{2}=\frac{\wp''(\omega_{j})
 \gamma}{\beta-e_{j}\delta},\qquad
\Hat{\Hat{a}}_{2}=3\frac{4e_{j}^{\,2}\beta^{2}+g_{3}\beta\delta
 -(4e_{j}^{\,3}+g_{3})e_{j}\delta^{2}}{2e_{j}(\beta-e_{j}\delta)^{2}}.
\end{split}
\label{eq:p125}
\end{align}
\textit{Functions}:
\begin{align}
\begin{split}
&z(x)=\wp(x),\qquad W(x)=\hW(x)=\Hat{\hW}(x)=\frac{\wp(x)^{2}-2e_{i}
 \wp(x)-e_{i}^{\,2}-e_{j}e_{k}}{\wp'(x)},\\
&E(x)=\frac{\wp''(x)}{\wp'(x)},\quad
 \hE(x)=\frac{\wp''(x)}{\wp'(x)}-\frac{2\wp'(x)}{\wp(x)-e_{i}},
 \quad\Hat{\hE}(x)=\frac{\wp''(x)}{\wp'(x)}-\frac{2\wp'(x)
 }{\wp(x)-e_{j}},
\\\end{split}
\end{align}
\textit{Potentials}:
\begin{subequations}
\begin{align}
V^{-}(x)=&\:\frac{b_{1}\wp(x)+\bar{b}_{0}}{2\wp'(x)^{2}}
 \left[b_{1}\wp(x)+\bar{b}_{0}-\frac{10}{3}\wp''(x)\right]
 -\frac{8}{9}\wp(x)+\frac{91}{18}\wp(2x)+\frac{3b_{1}}{2}-R,\\
V^{\rmi1}(x)=&\:\frac{b_{1}\wp(x)+\bar{b}_{0}}{2\wp'(x)^{2}}
 \left[b_{1}\wp(x)+\bar{b}_{0}-\frac{4}{3}\wp''(x)\right]
 -\frac{2}{9}\wp(x)+\frac{7}{18}\wp(2x)\notag\\
&+\frac{2\wp''(\omega_{j})}{\wp(x)-e_{j}}+4e_{j}+\frac{b_{1}}{2}
 -R,\\
V^{\rmj1}(x)=&\:\frac{b_{1}\wp(x)+\bar{b}_{0}}{2\wp'(x)^{2}}
 \left[b_{1}\wp(x)+\bar{b}_{0}+\frac{2}{3}\wp''(x)\right]
 +\frac{40}{9}\wp(x)-\frac{5}{18}\wp(2x)-\frac{b_{1}}{2}-R,\\
V^{\rmj2}(x)=&\:\frac{b_{1}\wp(x)+\bar{b}_{0}}{2\wp'(x)^{2}}
 \left[b_{1}\wp(x)+\bar{b}_{0}+\frac{2}{3}\wp''(x)\right]
 +\frac{4}{9}\wp(x)-\frac{5}{18}\wp(2x)\notag\\
&+\frac{2\wp''(\omega_{i})}{\wp(x)-e_{i}}+4e_{i}-\frac{b_{1}}{2}
 -R,\\
V^{+}(x)=&\:\frac{b_{1}\wp(x)+\bar{b}_{0}}{2\wp'(x)^{2}}
 \left[b_{1}\wp(x)+\bar{b}_{0}+\frac{8}{3}\wp''(x)\right]
 +\frac{10}{9}\wp(x)+\frac{55}{18}\wp(2x)-\frac{3b_{1}}{2}-R.
\end{align}
\end{subequations}
With the values of $b_{1}$ and $b_{0}$ in (\ref{eq:p125}), the first
term of each the potential can be expressed solely in terms of
$\wp(x)$ as
\begin{align}
\lefteqn{
\frac{b_{1}\wp(x)+\bar{b}_{0}}{2\wp'(x)^{2}}\left[b_{1}\wp(x)
 +\bar{b}_{0}+C\wp''(x)\right]}\hspace{30pt}\notag\\
&=\frac{[3e_{i}\wp(x)+e_{i}^{\,2}+2e_{j}e_{k}]^{2}}{18\prod_{l=1}^{3}
 (\wp(x)-e_{l})}+\frac{C}{6}\sum_{l=1}^{3}\frac{3e_{i}\wp(x)
 +e_{i}^{\,2} +2e_{j}e_{k}}{\wp(x)-e_{l}}.
\end{align}

\section{Parasupersymmetry and Generalized Superalgebras}
\label{sec:qpara}

In the case of $\cN=2$, it was shown~\cite{BT09} that any type A $2$-fold
SUSY system which has an intermediate Hamiltonian admits a realization
of second-order paraSUSY~\cite{RS88} and a generalized $2$-fold
superalgebra~\cite{Ta07a}. Thus, it is natural to ask whether an analogous
realization is possible in the present $\cN=3$ case. In what follows, we
show that it is indeed the case. More precisely, a type A $3$-fold SUSY
system of Class $(1,1)$ admits a realization of third-order
paraSUSY~\cite{To92,Kh92} and a generalized $3$-fold superalgebra found
in Ref.~\cite{Ta07c} while one of Class $(0,1)$ does only a restricted
version of the latter algebra. We shall first discuss the former
realization and then the latter.

\subsection{Parasupersymmetry of Order $3$ in Class $(1,1)$}

Higher-order paraSUSY was introduced in Refs.~\cite{To92,Kh92} as
a generalization of second-order one~\cite{RS88}. In the case of
third-order, it is characterized by the following algebraic relations:
\begin{subequations}
\label{eqs:pSS3}
\begin{align}
&(\bQ_{\rmP}^{\pm})^{3}\neq0,\qquad(\bQ_{\rmP}^{\pm})^{4}=0,\qquad
 \bigl[\bQ_{\rmP}^{\pm},\bH_{\!\rmP}\bigr]=0,
\label{eq:pSS31}\\
&(\bQ_{\rmP}^{\pm})^{3}\bQ_{\rmP}^{\mp}+(\bQ_{\rmP}^{\pm})^{2}
 \bQ_{\rmP}^{\mp}\bQ_{\rmP}^{\pm}+\bQ_{\rmP}^{\pm}\bQ_{\rmP}^{\mp}
 (\bQ_{\rmP}^{\pm})^{2}+\bQ_{\rmP}^{\mp}(\bQ_{\rmP}^{\pm})^{3}=
 6(\bQ_{\rmP}^{\pm})^{2}\bH_{\!\rmP}.
\label{eq:pSS32}
\end{align}
\end{subequations}
By the introduction of parafermionic coordinates $\psi_{\rmP}^{\pm}$
of order $3$ satisfying~\cite{Ta07c}
\begin{align}
(\psi_{\rmP}^{\pm})^{4}=0,\quad\bigl\{\psi_{\rmP}^{-},\psi_{\rmP}^{+}
 \bigr\}+\bigl\{(\psi_{\rmP}^{-})^{3},(\psi_{\rmP}^{+})^{3}\bigr\}=2I,
 \quad\bigl\{(\psi_{\rmP}^{-})^{2},(\psi_{\rmP}^{+})^{2}\bigr\}=I,
\end{align}
a quantum mechanical realization of paraSUSY of order $3$ is achieved
by defining the triple $(\bH_{\!\rmP},\bQ_{\rmP}^{\pm})$ as
\begin{subequations}
\label{eqs:QMrep}
\begin{align}
\bH_{\!\rmP}=&\:H_{0}(\psi_{\rmP}^{-})^{3}(\psi_{\rmP}^{+})^{3}+H_{1}
 (\psi_{\rmP}^{+}\psi_{\rmP}^{-}-(\psi_{\rmP}^{+})^{2}(\psi_{\rmP}^{-}
 )^{2})\notag\\
&+H_{2}((\psi_{\rmP}^{+})^{2}(\psi_{\rmP}^{-})^{2}-(\psi_{\rmP}^{+}
 )^{3}(\psi_{\rmP}^{-})^{3})+H_{3}(\psi_{\rmP}^{+})^{3}(\psi_{\rmP}^{-}
 )^{3},\\
\bQ_{\rmP}^{-}=&\:Q_{1}^{-}(\psi_{\rmP}^{-})^{3}(\psi_{\rmP}^{+})^{2}
 +Q_{2}^{-}(\psi_{\rmP}^{+}(\psi_{\rmP}^{-})^{2}-(\psi_{\rmP}^{+})^{2}
 (\psi_{\rmP}^{-})^{3})+Q_{3}^{-}(\psi_{\rmP}^{+})^{2}(\psi_{\rmP}^{-}
 )^{3},\\
\bQ_{\rmP}^{+}=&\:Q_{1}^{+}(\psi_{\rmP}^{-})^{2}(\psi_{\rmP}^{+})^{3}
 +Q_{2}^{+}(\psi_{\rmP}^{-}(\psi_{\rmP}^{+})^{2}-(\psi_{\rmP}^{-})^{2}
 (\psi_{\rmP}^{+})^{3})+Q_{3}^{+}(\psi_{\rmP}^{+})^{3}(\psi_{\rmP}^{-}
 )^{2},
\end{align}
\end{subequations}
where
\begin{align}
H_{k}=-\frac{1}{2}\frac{\rmd^{2}}{\rmd x^{2}}+V_{k}(x),\qquad
 Q_{k}^{\pm}=\pm\frac{\rmd}{\rmd x}+W_{k}(x).
\label{eq:QMrep4}
\end{align}
The linear space in which the system $(\bH_{\!\rmP},\bQ_{\rmP}^{\pm})$
shall be considered is $\fF\times\sV_{3}$ where $\fF$ is a linear
space of complex functions such as $L^{2}(\bbR)$ and $\sV_{3}=
\sum_{k=0}^{3}\sV_{3}^{(k)}$ is the parafermionic Fock space of order
$3$ composed of the $k$-parafermionic subspaces $\sV_{3}^{(k)}$
($k=0,\dots,3$).
Then, the latter system satisfies the third-order paraSUSY algebra
(\ref{eqs:pSS3}) if and only if \cite{Kh92,Kh93,Ta07c}
\begin{subequations}
\label{eqs:conp3}
\begin{align}
2H_{0}&=Q_{1}^{-}Q_{1}^{+}-2R_{1},\quad 2H_{1}=Q_{1}^{+}Q_{1}^{-}
 -2R_{1}=Q_{2}^{-}Q_{2}^{+}-2R_{2},\\
2H_{2}&=Q_{2}^{+}Q_{2}^{-}-2R_{2}=Q_{3}^{-}Q_{3}^{+}-2R_{3},\quad
 2H_{3}=Q_{3}^{+}Q_{3}^{-}-2R_{3},
\end{align}
\end{subequations}
where $R_{k}$ ($k=1,2,3$) are constants satisfying
\begin{align}
R_{1}+R_{2}+R_{3}=0.
\label{eq:R123}
\end{align}
Comparing now the paraSUSY conditions (\ref{eqs:conp3}) with the Class
$(1,1)$ conditions (\ref{eqs:cond5})--(\ref{eqs:cond7}), we immediately
notice that any type A $3$-fold SUSY system which belongs to Class
$(1,1)$ admits a realization of paraSUSY of order $3$ by the following
identifications:
\begin{align}
Q_{k}^{\pm}=P_{3\,4-k}^{\mp},\quad R_{k}=-C_{3\,4-k},\quad
 H_{0}=H^{-},\quad H_{1}=H^{\rmi1},\quad H_{2}=H^{\rmj1},\quad
 H_{3}=H^{+}.
\label{eq:A3pS3}
\end{align}
By the formulas (\ref{eq:C3i}) and the second relation in the above,
the constraint (\ref{eq:R123}) is expressed in terms of the type A
$\cN$-fold SUSY parameters as $R=0$, which is identical to the
constraint in the $\cN=2$ case (cf., Ref.~\cite{BT09}, Section~4).

The realization of third-order paraSUSY via the formulas (\ref{eqs:conp3})
admits another nonlinear relation, Ref.~\cite{Ta07c}, Eq.~(4.33). For the
present system, it reads by the formulas (\ref{eq:C3i}) and the second
relation in (\ref{eq:A3pS3})
\begin{multline}
(\bQ_{\rmP}^{-})^{3}(\bQ_{\rmP}^{+})^{3}+\left\{
 \begin{array}{c}
 \bQ_{\rmP}^{+}(\bQ_{\rmP}^{-})^{3}(\bQ_{\rmP}^{+})^{2}\\
 (\bQ_{\rmP}^{-})^{2}(\bQ_{\rmP}^{+})^{3}\bQ_{\rmP}^{-}
 \end{array}\right\}+\left\{
 \begin{array}{c}
 \bQ_{\rmP}^{-}(\bQ_{\rmP}^{+})^{3}(\bQ_{\rmP}^{-})^{2}\\
 (\bQ_{\rmP}^{+})^{2}(\bQ_{\rmP}^{-})^{3}\bQ_{\rmP}^{+}
 \end{array}\right\}+(\bQ_{\rmP}^{+})^{3}(\bQ_{\rmP}^{-})^{3}\\
=8\biggl(\Bigl(\bH_{\!\rmP}+R+\frac{a_{2}}{3}\Bigr)^{2}-b_{1}^{\,2}
 \biggr)\biggl(\bH_{\!\rmP}+R-\frac{2a_{2}}{3}\biggr).
\label{eq:pS3nl}
\end{multline}
We note that this algebraic relation holds irrespective of the
paraSUSY constraint (\ref{eq:R123}). Thus, in this sense the latter
algebra (\ref{eq:pS3nl}) is more general than the paraSUSY algebra
(\ref{eq:pSS32}).

In the subsector with the parafermion number $0$ and $3$, the
nonlinear algebra (\ref{eq:pS3nl}) reduces to
\begin{align}
\bigl\{(\bQ_{\rmP}^{-})^{3},(\bQ_{\rmP}^{+})^{3}\bigr\}
=&\:8\biggl(\Bigl(\bH_{\!\rmP}+R+\frac{a_{2}}{3}\Bigr)^{2}
 -b_{1}^{\,2}\biggr)\biggl(\bH_{\!\rmP}+R-\frac{2a_{2}}{3}\biggr)
 \biggr|_{\fF\times(\sV_{3}^{(0)}+\sV_{3}^{(3)})}.
\end{align}
This, together with the relations
\begin{align}
\bigl[(\bQ_{\rmP}^{\pm})^{3},\bH_{\!\rmP}\bigr]=
 \bigl\{(\bQ_{\rmP}^{\pm})^{3},(\bQ_{\rmP}^{\pm})^{3}\bigr\}=0,
\end{align}
which follow directly from the third-order paraSUSY relations in
(\ref{eq:pSS31}) forms a $3$-fold superalgebra. We can now easily
check that the latter algebra exactly coincides with type A $3$-fold
superalgebra (\ref{eqs:A3alg}) in Class $(1,1)$ by the conditions
(\ref{eq:c2}). Hence, an arbitrary type A $3$-fold SUSY system which
belongs to Class $(1,1)$ admits a realization of paraSUSY of order $3$
and the generalized type A $3$-fold superalgebra (\ref{eq:pS3nl}).

\subsection{Generalized $3$-fold Superalgebra in Class $(0,1)$ and Class
$(1,0)$}

Contrary to the case of Class $(1,1)$, any system belonging to Class
$(0,1)$ and $(1,0)$ admits neither paraSUSY of order $3$ nor
quasi-paraSUSY of order $(3,q)$~\cite{Ta07c}. The reason is the lack
of a `shape-invariant' condition at the place where an intermediate
Hamiltonian is absent. However, as we shall show shortly, a restricted
version of the generalized type A $3$-fold superalgebra (\ref{eq:pS3nl})
still holds in each of Class $(0,1)$ and Class $(1,0)$ with the same
parafermionic setting as (\ref{eqs:QMrep}) and (\ref{eq:A3pS3}). Indeed,
substituting the relations (\ref{eq:A3pS3}) into the formulas
(4.21)--(4.26) in Ref.~\cite{Ta07c} and using the intertwining
relations (\ref{eqs:caseI}) and the formula (\ref{eq:ant01}) for Class
$(0,1)$, and (\ref{eqs:caseI'}) and (\ref{eq:ant10}) for Class $(1,0)$,
respectively, we see that in the case of Class $(0,1)$ the following
algebraic relation holds in the subsector with the parafermion number
$0$, $2$, and $3$
\begin{multline}
(\bQ_{\rmP}^{-})^{3}(\bQ_{\rmP}^{+})^{3}+\left\{
 \begin{array}{c}
 \bQ_{\rmP}^{-}(\bQ_{\rmP}^{+})^{3}(\bQ_{\rmP}^{-})^{2}\\
 (\bQ_{\rmP}^{+})^{2}(\bQ_{\rmP}^{-})^{3}\bQ_{\rmP}^{+}
 \end{array}\right\}+(\bQ_{\rmP}^{+})^{3}(\bQ_{\rmP}^{-})^{3}\\
=8S_{2}^{(0,1)}(\bH_{\!\rmP}+R)
 \biggl(\bH_{\!\rmP}+R+\frac{a_{2}}{3}+b_{1}\biggr)
 \biggr|_{\fF\times(\sV_{3}^{(0)}+\sV_{3}^{(2)}+\sV_{3}^{(3)})}.
\label{eq:pnl01}
\end{multline}
and that in the case of Class $(1,0)$ the algebra which holds in the
subsector with the parafermion number $0$, $1$, and $3$ reads
\begin{multline}
(\bQ_{\rmP}^{-})^{3}(\bQ_{\rmP}^{+})^{3}+\left\{
 \begin{array}{c}
 \bQ_{\rmP}^{+}(\bQ_{\rmP}^{-})^{3}(\bQ_{\rmP}^{+})^{2}\\
 (\bQ_{\rmP}^{-})^{2}(\bQ_{\rmP}^{+})^{3}\bQ_{\rmP}^{-}
 \end{array}\right\}+(\bQ_{\rmP}^{+})^{3}(\bQ_{\rmP}^{-})^{3}\\
=8S_{2}^{(1,0)}(\bH_{\!\rmP}+R)
 \biggl(\bH_{\!\rmP}+R+\frac{a_{2}}{3}-b_{1}\biggr)
 \biggr|_{\fF\times(\sV_{3}^{(0)}+\sV_{3}^{(1)}+\sV_{3}^{(3)})}.
\label{eq:pnl10}
\end{multline}
In the whole parafermionic vector space $\fF\times\sV_{3}$, however,
no algebraic relations like (\ref{eq:pS3nl}) hold for Class $(0,1)$
and Class $(1,0)$. In fact, for Class $(0,1)$ each of the second term
in the l.h.s.\ of (\ref{eq:pS3nl}) is calculated in the subsector
with the parafermion number $1$ as
\begin{align}
\begin{split}
\bQ_{\rmP}^{+}(\bQ_{\rmP}^{-})^{3}(\bQ_{\rmP}^{+})^{2}&
 =2P_{33}^{-}P_{33}^{+}P_{32}^{+}(H^{\rmj1}-C_{31})P_{32}^{-}
 \bigr|_{\fF\times\sV_{3}^{(1)}},\\
(\bQ_{\rmP}^{-})^{2}(\bQ_{\rmP}^{+})^{3}\bQ_{\rmP}^{-}&
 =2P_{32}^{+}(H^{\rmj1}-C_{31})P_{32}^{-}P_{33}^{-}P_{33}^{+}
 \bigr|_{\fF\times\sV_{3}^{(1)}},
\end{split}
\end{align}
and cannot be expressed as a polynomial of $H^{\rmj1}$. Similarly,
for Class $(1,0)$ each of the third term in the l.h.s.\ of
(\ref{eq:pS3nl}) is calculated in the subsector with the parafermion
number $2$ as
\begin{align}
\begin{split}
\bQ_{\rmP}^{-}(\bQ_{\rmP}^{+})^{3}(\bQ_{\rmP}^{-})^{2}&
 =2P_{31}^{+}P_{31}^{-}P_{32}^{-}(H^{\rmi1}-C_{33})P_{32}^{+}
 \bigr|_{\fF\times\sV_{3}^{(2)}},\\
(\bQ_{\rmP}^{+})^{2}(\bQ_{\rmP}^{-})^{3}\bQ_{\rmP}^{+}&
 =2P_{32}^{-}(H^{\rmi1}-C_{33})P_{32}^{+}P_{31}^{+}P_{31}^{-}
 \bigr|_{\fF\times\sV_{3}^{(2)}},
\end{split}
\end{align}
and again cannot be expressed as a polynomial of $H^{\rmi1}$.

We note that both the nonlinear algebras (\ref{eq:pnl01}) and
(\ref{eq:pnl10}) are compatible with the type A $3$-fold superalgebra
(\ref{eqs:A3alg}). In fact, it is easy to check that both
(\ref{eq:pnl01}) and (\ref{eq:pnl10}) reduce to the anti-commutator
of the type A $3$-fold superalgebra (\ref{eq:A3alg2}) in the more
restricted subsector with the parafermion number $0$ and $3$ by
noting the condition (\ref{eq:c1}) for the former Class $(0,1)$ and
by (\ref{eq:c1'}) for the latter Class $(1,0)$.

\section{Discussion and Summary}
\label{sec:discus}

In this article, we have fully investigated the necessary and
sufficient conditions for a type A $3$-fold SUSY system to have
one or more sets of intermediate Hamiltonians and then made the
complete classification of them by the property of the $GL(2,\bbC)$
transformations. When only one set of intermediate Hamiltonians is
concerned, there are three different patterns in the existence and
called Class $(1,1)$, Class $(0,1)$, and Class $(1,0)$, respectively.
We have found that all the models which belong to Class $(1,1)$ are
not only solvable but also shape invariant while the ones which
belong to Class $(0,1)$ or Class $(1,0)$ are just quasi-solvable.
When more than one sets of intermediate Hamiltonians are concerned,
there emerge various patterns depending on the functional type of
each model. In Table~\ref{tb:pclas}, we summarize the possible
classes for each case of type A $3$-fold SUSY models.
\begin{table}
\begin{center}
\[
\tabcolsep=10pt
\begin{tabular}{ll}
\hline
 Case & Possible Classes\\
\hline
 I & $(1,1)$\\
 II & $(1,1)\supset(1,2)$\\
 II' & $(0,1)$\\
 III & $(1,1)$\\
 IV & $(1,1)\supset(1,2)\supset
  \begin{cases}(1,3)\\(1+1,1+1)\end{cases}$\\
 IV' & $(0,1)\supset\begin{cases}(0,2)\\(1;1)\end{cases}$\\
 V & $(0,1)\supset\begin{cases}(0,2)\supset(0,3)\\
  (1;1)\supset(1;2)\end{cases}$\\
\hline
\end{tabular}
\]
\caption{The possible classes of intermediate Hamiltonians for each
 case of type A 3-fold SUSY models.}
\label{tb:pclas}
\end{center}
\end{table}

It is now evident from Table~\ref{tb:pclas} that the structure of
higher-order intertwining operators is much richer than the degree
that one can classify them solely by the notion of reducibility
introduced in Refs.~\cite{AICD95,AIN95a}. It is not only because
the requirement of reality is restrictive but also because there
are various patterns in the existence of intermediate Hamiltonians.
Needless to say, the number of possible patterns drastically increase
as the order $\cN$ of intertwining operators gets higher.

Although we have not assumed in this article the reality of
Hamiltonians and thus have analyzed general complex Hamiltonians
by employing the $GL(2,\bbC)$ transformations, it is straightforward
to examine and classify real Hamiltonians by the use of the real
$GL(2,\bbR)$ transformations instead of $GL(2,\bbC)$. We only show
in Table~\ref{tb:rclas} an example of the real classification scheme
for that purpose. Note, however, that some of the possible classes
for Case IV and Case V in Table~\ref{tb:pclas} which can exist in
the complex case might be missing in the real case since the solutions
to the conditions (\ref{eq:con12}), (\ref{eq:con13}), (\ref{eq:con15}),
(\ref{eqs:con16}) or (\ref{eqs:con17}) are not necessarily real.
\begin{table}
\begin{center}
\[
\tabcolsep=10pt
\begin{tabular}{lll}
\hline
 Case & Class $(1,1)$ & Class $(0,1)$, $(1,0)$\\
\hline
 I & $a/2$ & \\
 II & $2z$ & \\
 II' & & $2z^{3}$\\
 III & $z^{2}/2$ & \\
 IV & $\pm2a(z^{2}-1)$ & \\
    & $\pm2a(z^{2}+1)$ & \\
 IV' & & $2az^{2}(z+1)$\\
 V & & $2z(1-z)(1-mz)$\\
   & & $z(z^{2}+2(1-2m)z+1)/2$\\
\hline
\end{tabular}
\]
\caption{A real classification scheme of type A 3-fold SUSY
 models with one set of intermediate Hamiltonians. In the above,
 $a\in\bbR$ is an arbitrary constant and $0<m<1$.}
\label{tb:rclas}
\end{center}
\end{table}\\

The realization of the variant generalized $3$-fold superalgebras
in Section~\ref{sec:qpara} indicates that the parafermionic formulation
like (\ref{eqs:QMrep}) could provide a more adequate and advantageous
framework to formulate $\cN$-fold SUSY than the conventional fermionic
formulation like (\ref{eq:ofrep}). In the conventional approach,
the type A $3$-fold superalgebra (\ref{eqs:A3alg}) cannot characterize
nor detect the existence of intermediate Hamiltonians at all. In
contrast to it, in the parafermionic approach on the one hand
the type A $3$-fold superalgebra is always realized in the subsector
with the fermion number $0$ and $3$, and on the other hand the
various patterns in the existence of intermediate Hamiltonians are
characterized by considering the other subsector with the fermion
number $1$ and/or $2$. In addition, generalized $\cN$-fold
superalgebra like (\ref{eq:pS3nl}) could provide an alternative
for defining paraSUSY of order $\cN$. In the present $\cN=3$ case,
the conventional defining algebra (\ref{eq:pSS32}) only produces
in essence the additional constraint (\ref{eq:R123}) whose physical
relevance is unclear. Hence, there seems, at least until now,
no physical evidence to claim that which definition is appropriate.

In principle, one can continue to study the $\cN>3$ cases, but as
already mentioned the number of different patterns in the existence
of intermediate Hamiltonians drastically increases as $\cN$ increases.
However, the number of different sets of intermediate Hamiltonians
would be limited to at most $4$ regardless of the number of patterns
due to the following reasons. First, the transformation formula
(\ref{eq:tfas}) which is valid for all $\cN\geq3$ could produce
algebraic equations of at most fourth-degree and thus at most four
additional sets of intermediate Hamiltonians would be admissible.
But type A $\cN$-fold SUSY systems for $\cN\geq3$ have at most
three independent free parameters $b_{i}$ ($i=0,1,2$) and thus
at most three different solutions among the four would be compatible
simultaneously. Hence, at most three \emph{additional} sets would be
available, which means that the maximum number of different sets
is four in total. By a similar argument we conclude that it would be
at most $3$ if one constraint on the parameters $b_{i}$ are inevitable
for the existence of one set of intermediate Hamiltonians, as in the
present $\cN=3$ case (\ref{eq:c2})--(\ref{eq:c1'}), since in this
case there are essentially at most two independent free parameters
and thus at most two algebraic solutions would be compatible
simultaneously.
The latter fact is indeed the reason why in the $\cN=3$ case there
are at most three different sets and thus are no classes such as
Class $(m,m+n)$ with $m+n>3$, Class $(m+l,m+n)$ with $m+l+n>3$,
Class $(0,n)$ with $n>3$, and Class $(l;n)$ with $l+n>3$.

\begin{acknowledgments}

T.~Tanaka's work was partially supported by the National Cheng Kung
University under the grant No.\ HUA:98-03-02-227.

\end{acknowledgments}



\bibliography{refsels}
\bibliographystyle{npb}



\end{document}